\documentclass[aps,prx,10pt,twocolumn,superscriptaddress,nofootinbib]{revtex4-2}
\usepackage[dvipsnames]{xcolor}
\usepackage{tikz,tikz-feynhand}
\usepackage{amsmath,amssymb,amsthm,bbm}
\usepackage[colorlinks,allcolors=Blue,linktocpage]{hyperref}
\usepackage{graphicx} 
\usepackage[normalem]{ulem}
\usepackage{placeins}
\usepackage{calrsfs}

\usepackage{CJK}
\usepackage[T1]{fontenc}
\usepackage[utf8]{inputenc}
\newcommand{\makeCJKtitle}{
    \begin{CJK*}{UTF8}{bkai}
        \maketitle
    \end{CJK*}}

\newcommand{\br}{\mathbf{r}}
\newcommand{\orb}{\mathrm{orb}}
\newcommand{\BI}{\mathbb{I}}
\newcommand{\rU}{\mathrm{U}}
\newcommand{\SU}{\mathrm{SU}}
\newcommand{\Sp}{\mathrm{Sp}}
\newcommand{\rO}{\mathrm{O}}
\newcommand{\SO}{\mathrm{SO}}
\newcommand{\rd}{\mathrm{d}}
\newcommand{\cC}{\mathcal{C}}
\newcommand{\tr}{\operatorname{tr}}
\newcommand{\Tr}{\operatorname{Tr}}
\newcommand{\Dlt}{\tilde{\Delta}}

\begin{document}

\title{\boldmath Generalizing Deconfined Criticality to 3D $N$-Flavor $\SU(2)$ Quantum Chromodynamics on the Fuzzy Sphere}

\author{Emilie Huffman}
\email{ehuffman@wfu.edu}
\affiliation{Department of Physics and Center for Functional Materials, 
Wake Forest University, Winston-Salem, North Carolina 27109, USA}
\author{Zheng Zhou (周正)}
\affiliation{Perimeter Institute for Theoretical Physics, Waterloo, Ontario N2L 2Y5, Canada}
\affiliation{Department of Physics and Astronomy, University of Waterloo, Waterloo, Ontario N2L 3G1, Canada}
\author{Yin-Chen He}
\affiliation{Perimeter Institute for Theoretical Physics, Waterloo, Ontario N2L 2Y5, Canada}
\affiliation{C.~N.~Yang Institute for Theoretical Physics, Stony Brook University, Stony Brook, NY 11794-3840}
\author{Johannes S. Hofmann}
\email{jhofmann@pks.mpg.de}
\affiliation{Max Planck Institute for the Physics of Complex Systems, N\"othnitzer Strasse 38, Dresden 01187, Germany}

\date{\today}

\begin{abstract}
    The infra-red behaviour of gauge theories coupled to matter remains an open problem in quantum field theory. For a given gauge group, such theories are expected to flow to an interacting conformal fixed point over a range of fermion or scalar flavours, known as the `conformal window.' Their nature is important for understanding critical phases and phase transitions beyond the Landau paradigm like the deconfined quantum critical point (DQCP), yet remains challenging for conventional non-perturbative approaches. In this work, we study a family of fuzzy-sphere models corresponding to non-linear sigma models with $\Sp(N)$ global symmetry extended to the strongly-coupled region. These theories are expected have an infra-red fixed point described by $\SU(2)$ quantum chromodynamics (QCD) in three space-time dimensions with $N$ flavours of fermions. They can be viewed as a generalisation of the $\SO(5)$ DQCP, corresponding to $N=2$. We investigate them using quantum Monte Carlo for $N$ up to $16$. We find evidence that for $N\geq4$ the phase diagram contains a critical phase that appears to be absent for $N=2$. Within this phase, we measure the two-point correlation function and the excitation spectrum, which exhibit emergent conformal symmetry. We also extract the scaling dimension $\Delta_\phi$ of a leading operator and find consistency with large-$N$ expectations.
\end{abstract}

\makeCJKtitle
\tableofcontents

\newcommand{\bQ}{\mathbf{Q}}
\newcommand{\bQt}{\tilde{\mathbf{Q}}}
\newcommand{\bq}{\mathbf{q}}
\newcommand{\bbZ}{\mathbb{Z}}
\newcommand{\rG}{\mathrm{G}}
\newcommand{\bg}{\mathbf{g}}
\newcommand{\bA}{\mathbf{A}}
\newcommand{\half}{\tfrac{1}{2}}
\newcommand{\psb}{\bar{\psi}}
\newcommand{\WZW}{\text{WZW}}
\newcommand{\cO}{\mathcal{O}}

\section{Introduction}

Critical phenomena account for one of the central topics in modern condensed matter physics~\cite{Cardy1996Scaling,Sachdev2011Quantum}. They appear widely in quantum phase transitions, \textit{i.\,e.}~singularities of ground state upon tuning a parameter in the Hamiltonian, exhibiting power-law correlations and scale invariance, and extend beyond physics to complex systems. The seemingly complicated phenomena can be governed by simple underlying theories, a powerful idea known as universality: distinct critical systems share the same universal numbers, \textit{i.\,e.}~critical exponents, regardless of microscopic details. At low energies, critical points often exhibit an emergent conformal symmetry --- they become invariant under transformations that locally preserve angles while changing shapes. Conformal field theory (CFT) captures the essence of the resulting universal behaviour~\cite{Polyakov1970Conformal,Cardy1996Scaling,Sachdev2011Quantum}, and produces many predictions. CFT also connects with high energy physics, providing insights into string theory~\cite{Polchinski1998String}, quantum gravity~\cite{Maldacena1998AdSCFT}, renormalisation group (RG) flow~\cite{Zamolodchikov1986Irreversibility}, \textit{etc.} Many CFTs are well understood in 2D~\cite{DiFrancesco1997CFT,Ginsparg1988CFT,Belavin1984BPZ}, but their solutions in 3D remain difficult, although some conformal data has been determined at high precision by conformal bootstrap~\cite{Poland2018Bootstrap,Rychkov2023Bootstrap}. 

One fertile ground for discovering exotic 3D CFTs is phase transitions beyond Landau's paradigm of spontaneous symmetry breaking. A pioneering example of transitions beyond Landau is the deconfined quantum critical point (DQCP) originally proposed as direct continuous transition between the N\'eel antiferromagnet (AFM) and a valence-bond solid (VBS) phase~\cite{Senthil2003DQCP,Motrunich2003DQCP,Senthil2004DQCP,Read1990DQCP,Read1989DQCP,Murthy1989DQCP} (Also see a recent review~\cite{Senthil:2023vqd}). It also features several interesting properties, including an enhanced $\SO(5)$ symmetry at the transition point~\cite{Nahum2015DQCP}. Several field theory descriptions in terms of dual critical gauge theories have been proposed, such as 3D quantum electrodynamics (QED$_3$) with a $\rU(1)$ gauge field coupled to 2 complex scalars~\cite{Senthil2003DQCP} or the 3D quantum chromodynamics (QCD$_3$) with a $\SU(2)$ gauge field coupled to 2 fundamental fermions~\cite{Wang2017DQCP}. However, numerical studies have shown that the DQCP is weakly first order~\cite{Nahum2015DQCP,Zhou2023SO5,Takahashi:2024xxd}, plausibly being pseudo-critical, with approximate conformal symmetry plausibly emerging from nearby complex fixed points~\cite{Wang2017DQCP,Gorbenko:2018ncu,Gorbenko:2018dtm}. A natural question is therefore whether this set-up can be generalized to allow a continuous phase transition, described by a genuine CFT. One such attempt has been the transition between the $\SU(N)$ AFM and a VBS~\cite{Lou2009SUN,Kaul2011SUN,Kaul2012SUN,Block2013SUN} phase, which is shown to be continuous at large enough $N$, and described by scalar QED$_3$ with $N$ scalars (\textit{i.\,e.}~$\mathbb{C}\mathrm{P}^{N-1}$ theory). With the help of the field theory descriptions, we can consider another candidate: the $\SU(2)$ QCD$_3$ with $N$ fermions with a $\Sp(N)/\bbZ_2$ global symmetry; the original AFM-VBS DQCP corresponds to the $N=2$ case. 

On the other hand, the nature of gauge theories has been a long-standing problem in the study of quantum fields. Gauge theories provide the successful theoretical framework for the dynamics of elementary particles. Furthermore, coupling matter to gauge fields is one of a few known ways to construct interacting critical theories. For a given gauge group, the RG flow lands on an interacting infrared CFT only at a certain range of flavour number $N$ of the matter coupled to the gauge field instead of confinement or spontaneous symmetry breaking~\cite{Banks1981Window}. The determination of this range, known as the conformal window, is an outstanding problem in quantum field theory. In four dimensions, the conformal window is bounded from above and below $N_{c1}<N<N_{c2}$. In 3D, the gauge theory is conformal when the number of flavour exceeds a critical value $N>N_c$. The conformal window of the QED$_3$ has been extensively studied by lattice-based method~\cite{Hands:2002dv,Hands:2004bh,Raviv:2014xna,Karthik:2016bmf,Karthik:2015sgq,Karthik:2015sza} and dimension and large-flavour-number expansion~\cite{Appelquist:1985vf,Appelquist:1986fd,Appelquist:1986qw,Appelquist:1988sr,Gracey:1993iu,Gracey:1993sn,DiPietro:2015taa,Giombi:2015haa,Kotikov:2016prf,Gusynin:2016som,DiPietro:2017kcd,Herbut:2016ide}; the lattice calculation of $\SU(2)$ QCD$_3$~\cite{Dagotto:1991br,Karthik:2018nzf} shows that $N_c$ lies between 4 and 6; $\SU(N)$ QCD$_3$ at large $N_c$ has also been studied~\cite{Karthik:2016bmf}.

The fuzzy-sphere regularisation~\cite{Zhu2022} has emerged as a new powerful method to study 3D CFTs. By studying interacting quantum systems on the fuzzy (non-commutative) sphere, the method realises $(2+1)$D quantum phase transitions on the geometry $S^2\times\mathbb{R}$. Compared with conventional lattice simulations, this is a regularization scheme in the continuum~\cite{Ippoliti:2018ojo} that preserves the rotation symmetry exactly, which allows us to observe the emergent conformal symmetry directly and extract the conformal data efficiently. The power of this approach has been demonstrated in the context of the 3D Ising transition~\cite{Hu2023Mar,Han2023Jun,Hu2024,Fardelli2024,Fan2024,Dong2025,Hofmann2023,Voinea2024,Laeuchli2025,Wiese2025,Hao2026}, establishing the emergent conformal symmetry and calculating a wealth of conformal data at high precision. This approach has since been applied to various defects, boundaries~\cite{Hu2023Defect,Zhou2024Jan,Cuomo2024,Zhou2024Jul,Dedushenko2024} and various other CFTs~\cite{Han2023Dec,Zhou2023SO5,Zhou2024Oct,Yang2025Jan,Fan2025,ArguelloCruz2025,EliasMiro2025,He2025Jun,Taylor2025,Yang2025Jul,Zhou2025Jul,Zhou2025Sep,Voinea2025,Dey2025,Guo2025,Tang2025,Chen2024DQCP}.

Notably, the $\SO(5)$ DQCP has been studied on a fuzzy-sphere set-up with $N_f=4$ flavours of fermions with a $\Sp(2)/\bbZ_2$ flavour symmetry at half filling~\cite{Zhou2023SO5,Ippoliti:2018ojo,Wang:2020xza}. The correspondence with DQCP can be seen through matching the anomaly of the non-linear sigma model (NLSM) on the Grassmannian $\Sp(2)/(\Sp(1)\times\Sp(1))\equiv S^4$ with a level-1 Wess-Zumino-Witten (WZW) term~\cite{Nahum2015DQCP,Lee2014WZW}. A natural generalisation would be $N_f=2N$ flavours of fermions with $\Sp(N)$\footnote{Here we adopt a notation that $\Sp(N)$ denotes the group of $2N\times2N$ unitary symplectic matrices.} symmetry at half filling and even $N$. This set-up realises the NLSM-WZW on
\begin{equation*}
    \frac{\Sp(N)}{\Sp(N/2)\times\Sp(N/2)}
\end{equation*}
extended to the strongly-coupled region, which matches with the $N$-flavour $\SU(2)$ QCD$_3$~\cite{Zhou2024Oct}. A small value $N=4$ has been studied through exact diagonalisation (ED) and hint of conformal symmetry has been observed. However, due to the exponentially divergent computational cost, larger $N$ are inaccessible through ED and DMRG. 

These models are well-suited for auxiliary-field quantum Monte Carlo (QMC) sampling, however. As we will demonstrate below in more detail, the effective Hamiltonian for any given auxiliary-field configuration is manifestly $\SU(N) \subset \Sp(N)$-symmetric for even $N$ and therefore block-diagonal.\footnote{Note that some symmetries of the Hamiltonian may be broken for a given auxiliary field configuration and will only be restored upon averaging over the configuration space.} This implies that, for example, the weight of a configuration takes the form $\det(M)^{N}$, where $M$ represents the fermion determinant of a flavor-pair. Consequently, the computational effort is essentially independent of $N$. Furthermore, a particle-hole symmetry at half-filling guarantees the absence of the notorious fermion sign problem, allowing large-scale QMC simulations \cite{Ippoliti:2018ojo,PhysRevB.71.155115}.

In this work, we use such QMC simulations in the fuzzy-sphere set-up to study the $\Sp(N)$-symmetric models at arbitrary $N$ with a particular focus on the conformal window and the associated possibility of generalized deconfined criticality. We first give theoretical details about the fuzzy-sphere model and match it with the $\SU(2)$ QCD$_3$ through NLSM-WZW, as well as observables of interest. Then we present the numerical results for the phase diagram, the conformal correlators, the scaling dimension $\Delta_\phi$ of a leading operator, and the state-operator correspondence. We conclude with a discussion on the implications and prospective extensions of our work.

\section{Model and Method}
\label{model}

\subsection{Fuzzy-Sphere Model}

To construct the NLSM on the Grassmannian $\Sp(N)/(\Sp(N/2)\times\Sp(N/2))$ with the fuzzy-sphere regularization, we start with $N_f=2N$ flavours of fermions moving on a sphere with a $4\pi s$-monopole at its center. Due to the presence of the monopole, the single-particle eigenstates form highly degenerate quantised Landau levels. The single-particle (non-interacting) ground state, \textit{i.\,e.}, the lowest Landau level (LLL), has a degeneracy $N_\orb=2s+1$ for each flavour. We partially fill the LLL and set the gap between the LLL and higher Landau levels to be much larger than any other energy scale in the system. In this case, we can effectively project the system into the LLL. After the projection, the fermion operator $\psi^i(\br)$ ($i=1,\dots,N_f$) can be expressed in terms of the annihilation
operators on the LLL
\begin{equation}
    \psi_i^\dagger(\br)=\frac{1}{R}\sum_{m=-s}^s Y_{sm}^{(s)}(\br)c_{mi}^\dagger
\end{equation}
where $Y_{sm}^{(s)}(\br)$ are the monopole spherical harmonics, $m=-s,\dots,s$ labels the Landau orbitals, and the radius of the sphere is taken as $R=N_\orb^{1/2}$. 

The model possesses a maximal global symmetry of $\SU(2N)$ alongside the $\rU(1)$ charge conservation that decouples at the critical point. To construct an interaction Hamiltonian that breaks the global symmetry from $\SU(2N)$ to $\Sp(N)$, we consider the $\Sp(N)$-invariant fermion bilinears:
\begin{enumerate}
    \item the fermion density $n(\br)=\psi_i^\dagger(\br)\psi^i(\br)$ which is $\SU(2N)$-invariant, and
    \item the pairing operator $\Delta(\br)=\psi^i(\br)\Omega_{ij}\psi^j(\br)/2$, where $\Omega=\BI_{N}\otimes i\sigma^y$, which is $\Sp(N)$-invariant but not $\SU(2N)$-invariant. 
\end{enumerate}
The Hamiltonian consists of the local density-density interaction and pair-pair interaction~\cite{Zhou2023SO5,Zhou2024Oct} 
\begin{equation}
    H=\int\rd^2\br\,\left(Un(\br)^2-\frac{V}{N}\Delta^\dagger(\br)\Delta(\br)\right).
    \label{eq:hmt}
\end{equation}
After the projection onto the LLL, we express the model in terms of the fermion operators $c_m^i,c^\dagger_{m,i}$, where $m$ labels the Landau orbital, through the angular components of $n$ and $\Delta$,
\begin{align}
    n(\br)&=\frac{1}{R^2}\sum_{l=0}^{2s}\sum_{m=-l}^ln_{lm}Y_{lm}(\br)\nonumber\\
    \Delta(\br)&=\frac{1}{R^2}\sum_{m=-2s}^{2s}\Delta_{2s,m}Y_{2s,m}^{(2s)}(\br)\nonumber\\
    H&=\frac{4\pi U}{R^2}\sum_{lm}n_{lm}^\dagger n_{lm}-\frac{4\pi V}{NR^2}\sum_m\Delta^\dagger_{2s,m}\Delta_{2s,m}. \label{eq:hamiltonian}
\end{align}
where the components $n_{lm}$ and $\Delta_{2s,m}$ take the form
\begin{align}
    n_{lm}&=\sum_{m_1m_2}\Lambda^{(n)}{}_{m_1m_2}^{lm}c^\dagger_{im_1}c^i_{m_2}\nonumber\\
    \Delta_{2s,m}&=\sum_{m_1m_2}\Lambda^{(\Delta)}{}_{m_1m_2}^{2s,m}c^i_{m_1}\Omega_{ij}c^j_{m_2}
    \label{eq:def_den}
\end{align}
in the orbital space, and we give the expression for coefficients $\Lambda^{(n)}{}_{m_1m_2}^{lm}$ and $\Lambda^{(\Delta)}{}_{m_1m_2}^{2s,m}$ in Appendix~\ref{app:den}. Apart from the $\SO(3)$ rotation symmetry and the $\Sp(N)$ global symmetry, this model also has an anti-unitary particle-hole symmetry
\begin{equation}
    \mathcal{P}:\quad c_{m}^j\mapsto c^\dagger_{mk}\Omega^{jk},\quad i\mapsto -i.
\end{equation}

In the following sections, we show that this model is most likely to realise the $\SU(2)$ QCD$_3$ with $N$ flavours of fundamental fermions. We do this in two steps: In Sec.~\ref{sec:nlsm}, we show that the fuzzy-sphere model can be described by a NLSM with WZW level-1; in Sec.~\ref{sec:cfwin}, we discuss the phase diagram of the NLSM, and we especially show that at large enough $N$, it contains a stable fixed point corresponding to the QCD. 

\subsection{Matching the WZW Level of NLSM}
\label{sec:nlsm}

In this section, we show that the model on the lowest Landau level matches the NLSM on the Grassmannian 
\begin{equation}
    \frac{\Sp(N)}{\Sp(N/2)\times\Sp(N/2)}
    \label{eq:grass}
\end{equation}
with WZW level-1, \textit{i.\,e.}, they have the same symmetry and anomaly. Its action is
\begin{align}
    S_\text{NLSM}[\bQ]&=\frac{1}{g}\int\rd^3x\,\tr( \partial^\mu\bQ )^2+k\,\Gamma_\WZW[\bQ]\nonumber\\
    \Gamma_\WZW[\bQ]&=\frac{2\pi i}{(16\pi)^2}\int_0^1\rd u\int\rd^3x\nonumber\\
    &\qquad\quad\times\epsilon^{\mu\nu\rho\sigma}\tr(\bQt\,\partial_\mu\bQt\,\partial_\nu\bQt\,\partial_\rho\bQt\,\partial_\sigma\bQt)
    \label{eq:act_nlsm}
\end{align}
where $g$ is the stiffness of the NLSM, $k$ is the WZW level, $x=(\tau,\br)$, and $\bQ(x)$ is a $2N\times 2N$ matrix field that parametrises the Grassmannian
\begin{equation}
    \bQ=\bg\Sigma\bg^{-1},\quad \bg\in\Sp(N),\quad\Sigma=\begin{pmatrix}
        \BI_{N}&0\\0&-\BI_{N}
    \end{pmatrix}.
\end{equation}
The WZW term is expressed by extending it into an auxiliary fourth dimension $\bQt(x,u)$ parametrised by $0\leq u\leq 1$ with boundary condition 
\begin{equation*}
    \bQt(x,u=1)=\bQ(x),\quad \bQt(x,u=0)=\Sigma
\end{equation*}

We shall derive the same action from the fuzzy-sphere model. The calculation for the specific case $N=2$ has already been performed by Lee and Sachdev~\cite{Lee2014WZW}. Here we sketch the process and leave the details to Appendix~\ref{app:wzw}. We start from free fermions on the lowest Landau level coupled to the matrix field $\bQ(x)$
\begin{equation}
    S[\psi,\psb,\bQ]=\int\rd^3x\, \bar\psi_i\frac{\partial\psi^i}{\partial\tau}+\frac{\tr(\partial^\mu\bQ)^2}{2g_0}-\lambda\psb_iQ^i{}_j \psi^j
\end{equation}
The fermion fields are projected to the lowest Landau level
\begin{equation*}
    \psi^i(x)=\sum_m\phi_m(\br)c^i_m(\tau)
\end{equation*}
where $m$ is the orbital index and $\phi_m(\br)$ is the single particle wavefunction of the lowest Landau level. Physically, the coupling between fermions to $\bQ$ captures the four-fermion interactions --- integrating out the $Q$ field results in an effective four-fermion term similar to Eq.~\eqref{eq:hmt} to the leading order, together with interactions with more fermions at higher order. 

Integrating out the fermions gives an effective action $S_\text{eff}[\bQ]$ of the matrix field $\bQ$. In the long wavelength limit, this effective action is equivalent to $S_\text{NLSM}$ in Eq.~\eqref{eq:act_nlsm} with WZW level $k=1$. This can be derived perturbatively by expanding $\bQ$ in vicinity of a reference configuration which we take as $\Sigma$
\begin{equation*}
    \bQ=\Sigma+\delta\bQ.
\end{equation*}
To the second order of $\delta\bQ$, $S_\mathrm{eff}$ gives the self-energy correction for $\bQ$; to the fourth order, the box diagram gives the WZW term that we desire
\begin{multline}
    \lim_{q\to 0}\begin{tikzpicture}[baseline=(bs.base)]
        \begin{feynhand}
            \vertex (bs) at (0,-0.1);
            \vertex(a) at (-0.6, 0.6);
            \vertex(b) at ( 0.6, 0.6);
            \vertex(c) at ( 0.6,-0.6);
            \vertex(d) at (-0.6,-0.6);
            \vertex (a1) at (-1,1) ;
            \vertex (b1) at (1,1)  ;
            \vertex (c1) at (1,-1) ;
            \vertex (d1) at (-1,-1);
            \propag[sca] (a1) to (a);
            \propag[sca] (b1) to (b);
            \propag[sca] (c1) to (c);
            \propag[sca] (d1) to (d);
            \propag[fermion] (a) to[edge label] (b);
            \propag[fermion] (b) to[edge label] (c);
            \propag[fermion] (c) to[edge label] (d);
            \propag[fermion] (d) to[edge label] (a);
        \end{feynhand}
    \end{tikzpicture}\\
    =\frac{3i}{64\pi}\int\rd^3x\,\tr(\Sigma\delta\bQ\,\partial_0\delta\bQ\,\partial_1\delta\bQ\,\partial_2\delta\bQ)\\=\Gamma_\WZW[\bQ],
\end{multline}
where the solid arrows and the dashed lines denote, respectively, the propagator of fermions and matrix fields. In Appendix~\ref{app:wzw}, we perform the detailed calculation on a large flat space with periodic boundary condition. We expect the result to extend to the LLL on the sphere, as the matching of the WZW level is not sensitive to the geometric details.

\subsection{Conformal Window of Gauge Theories}
\label{sec:cfwin}

Now that we have matched the fuzzy-sphere model and the NLSM-WZW, we discuss the phase diagram of the NLSM. We note that although the NLSM can only describe the spontaneous symmetry breaking (SSB) in the renormalisable region, by extending it to the strongly-coupled region, its phase diagram may contain critical gauge theories~\cite{Komargodski2017NLSM,Zou:2021dwv}. These gauge theories have the same symmetry and anomaly as the fuzzy-sphere model, and are thus natural end-points of RG-flow for the fuzzy-sphere model.

A natural candidate in the phase diagram is $\SU(2)$ QCD$_3$, \textit{i.~e.}~$\SU(2)$ gauge field coupled to $N$ flavours of fermions in the fundamental representation. To see this, we consider $N$ fermions $\psi^{ias}$ in the bi-fundamental representation of $\Sp(N)$ flavour symmetry and the $\SU(2)$ gauge group, where the three indices are respectively flavour, gauge and spinor.\footnote{The fermions here are different from the fermions in the microscopic model in Sec.~\ref{sec:nlsm}. The Lagrangian reads
\begin{equation*}
    \mathcal{L}[\psi^{ia},A_\mu]=\psi^{ias_1}\Omega_{ij}\Omega_{ab}\Omega_{s_1s'}i(\gamma^\mu)^{s'}{}_{s_2}(D_{A,\mu}\psi)^{jbs_2}
\end{equation*}
where the covariant derivative $D_{A,\mu}=\partial_\mu-iA_\mu$, the $\Sp(N)$ flavour index $i,j=1,\dots,2N$, the $\Sp(1)\cong \SU(2)$ gauge index $a,b=1,2$, and the $\Sp(1)\cong \SU(2)$ spinor index $s=1,2$. From this Lagrangian, the $\Sp(N)$ symmetry is explicit.} We then couple it to a bosonic mass field $\bQ$ in the $\Sp(N)$ Grassmannian through $\bar{\Psi}\bQ\Psi=\bar{\psi}_{ias}Q^i{}_j\psi^{jas}$. Condensing $\bQ$ gives the spontaneous symmetry broken phase described by the NLSM; on the other hand, when $\bQ$ is gapped, we recover the Lagrangian of the QCD.

\begin{figure}
    \centering
    \includegraphics[width=0.45\textwidth]{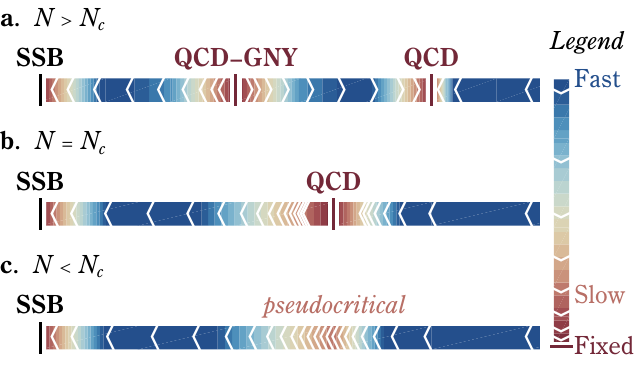}
    \caption{The putative RG-flow diagram of the NLSM extended to the strongly-coupled region. The colour and the intervals between arrows mark the rate of the RG flow. The red and black bars denote the conformal and non-conformal fixed points.}
    \label{fig:ph_diag}
\end{figure}

The phase diagram then depends on the fate of the QCD in the infrared. It is generally believed that $\SU(2)$ QCD$_3$ flows to a conformal fixed point at large enough $N$. In this range, the phase diagram contains three fixed points (Figure\ref{fig:ph_diag}a): a stable fixed point of the spontaneous symmetry-breaking phase, a stable conformal fixed point of the critical $\SU(2)$ QCD, and an unstable conformal fixed point corresponding to their phase transition, described by $\SU(2)$ QCD coupled to a Gross-Neveu-Yukawa field in the adjoint representation of the $\Sp(N)$. In the large-$N$ limit, the operator spectrum of the stable QCD fixed point has been studied using the standard perturbative computation technique~\cite{Xu2008LargeN}.\footnote{Note that $N$ in our convention is twice the $N$ in their convention.}
\begin{align}
    \Delta_{S^-}&=2+\frac{64}{3\pi^2}\frac{1}{N}+\mathcal{O}(N^{-2})\nonumber\\
    \Delta_\phi&=2-\frac{32}{3\pi^2}\frac{1}{N}+\mathcal{O}(N^{-2}).
    \label{largenres}
\end{align}
Specifically, the scaling dimension for the bilinear $S^-=\psi^i\psi^j\Omega_{ij}$ with odd parity in the singlet representation of $\Sp(N)$, and
\begin{equation}
    \phi^{[ij]}=\psi^{[i}\psi^{j]}-\text{(trace)}
    \label{eq:op_antisym}
\end{equation}
in the traceless antisymmetric rank-2 tensor representation of $\Sp(N)$, where the contraction of the other indices is omitted. Also note that this singlet, with odd parity, corresponds to the singlet fermion mass, while the parity-even singlet $S^+$ that controls the RG flow in Figure \ref{fig:ph_diag}a can be written as a four-fermion operator in the Lagrangian. 

As we decrease $N$, the two fixed points QCD and QCD-GNY approach each other, and at a critical $N_c$, they collide into one fixed point (Figure \ref{fig:ph_diag}b). This critical $N_c$ is signaled by an exactly marginal singlet $\Delta_{S^+}=3$ that controls the phase diagram. When we further decrease $N$ to $N<N_c$, the fixed point becomes two conjugate complex fixed points~(Figure \ref{fig:ph_diag}c), and the QCD flows directly to the SSB fixed point. The complex fixed points are described by complex CFTs. If they are sufficiently close to the real axis, the RG flow on the real axis in their vicinity slows down significantly. This is known as pseudo-criticality and is studied in deconfined criticality on the fuzzy sphere~\cite{Zhou2023SO5} which corresponds to $N=2$ here, as well as the 2D five-state Potts model~\cite{Gorbenko:2018ncu,Gorbenko:2018dtm}.

A similar scenario holds for 3D critical gauge theories involving $N$ fermions (or critical scalars) coupled to a more general dynamical gauge field (\textit{e.\,g.}~$\SU(k)$, $\rU(k)$, $\Sp(k)$). For each gauge group, there exists a critical value of flavour number $N_{c}$ above which the critical gauge theory flows to a conformal fixed point. The $N>N_{c}$ region where the 3D gauge theory flows to an interacting conformal fixed point is known as the conformal window. For QCD in 4D, there exist two critical $N_{c}$'s: $N<N_{c1}$ flows to SSB, $N_{c1}<N<N_{c2}$ flows to an interacting conformal fixed point, and $N>N_{c2}$ flows to a free fixed point. The conformal window problem, \textit{i.\,e.}, determining the exact value of the critical flavour number, has been an outstanding problem for condensed matter and high energy physicists. In 3D, the QCDs are interacting throughout $N>N_c$ even at large-$N$. In this paper, we shall focus on $N>N_c$ case, and examine the evidence for conformal symmetry at the stable QCD fixed point, as well as obtain its conformal data. 

\subsection{Method and Observables}

Here we give the general approach and some model-specific aspects to studying the $\Sp(N)$ symmetric NLSM through auxiliary-field quantum Monte Carlo for arbitrary $N$. To facilitate the Hubbard-Stratonovich decoupling, we write the Hamiltonian, Eq.~\eqref{eq:hamiltonian}, as a sum of perfect squares of hermitian operators,
\begin{multline}
    H = \frac{4\pi}{R^2}\frac{U}{2} \sum_{l,m}\left(e^{i\frac{\pi}{4}}n_{lm}^\dagger + e^{-i\frac{\pi}{4}}n_{lm}\right)^2 \\-\frac{4\pi}{R^2}\frac{V}{4N}\sum_m \left[\left( \Delta^\dag_{2s,m} + \Delta^{\phantom{\dag}}_{2s,m} \right)^2\right. \\
    + \left.\left( i \Delta^\dag_{2s,m} - i \Delta_{2s,m} \right)^2\right].
\end{multline}

From this point on, we fix $U=1$ as the unit of energy. We can next readily introduce auxiliary fields, $\eta^z_{lm,\tau}$, $\eta^+_{m,\tau}$, and $\eta^-_{m,\tau}$, for three kinds of perfect squares, respectively, on every imaginary-time slice $\tau$ in the trotterized partition sum $Z=\Tr{\prod_\tau e^{-\Delta \tau H}}$. The attentive reader may have noticed that the fermion bilinears coupling to $\eta^+_{m,\tau}$ and $\eta^-_{m,\tau}$ do not preserve the fermion particle number. Hence, a partial particle-hole transformation, $c^j_m \mapsto c^\dag_{mj} $ for $j = N+1,\dots,2N$, is performed to restore the fermion particle number in this new `computational' basis, $\tilde{c}^{j,\uparrow}_m=c^j_m$ and $\tilde{c}^{j,\downarrow}_m=c^{\dag}_{m,j+N}$ with $j=1,\cdots, N$. Eq.~\eqref{eq:def_den} is expressed in the new basis as
\begin{align}
    n_{lm} &= \sum_{m_1m_2} \Lambda^{(n)}{}^{lm}_{m_1m_2} 
    \left( 
    \tilde{c}^{\dag}_{m_1,j,\uparrow} \tilde{c}^{j,\uparrow}_{m_2} - 
    \tilde{c}^{\dag}_{m_2,j,\downarrow} \tilde{c}^{j,\downarrow}_{m_1}
    \right)\nonumber\\
    \Delta_{2s,m} &= \sum_{m_1m_2} \Lambda^{(\Delta)}{}^{2s,m}_{m_1,m_2}
    \tilde{c}^{\dag}_{m_2,j,\downarrow} \tilde{c}^{j,\uparrow}_{m_1} \nonumber\\
    \Delta^\dag_{2s,m} &= \sum_{m_1m_2} \Lambda^{(\Delta)}{}^{2s,m}_{m_1,m_2}
    \tilde{c}^{\dag}_{m_1,j,\uparrow} \tilde{c}^{j,\downarrow}_{m_2} \,.
\end{align}

For any given auxiliary-field configuration, the partition sum is $\SU(N)$ symmetric\footnote{Note that the form factors are independent of the flavour index.} and therefore exhibits $N$ identical $2N_\orb\times2N_\orb$ blocks. In a typical model with short-range interaction, the numerical effort of the DQMC algorithm scales cubically with the number of fermions per block. However, this is assuming that the auxiliary fields couple to local operators, or more precisely that the rank of those operators is independent of system size and typically rather small. In the fuzzy-sphere model, the form factors $\Lambda$ generically exhibit $N_\orb$ non-zero eigenvalues such that their rank scales with the system size as well. Consequently, the numerical effort here scales as $(2N_\orb)^4L_{\text{trot}}$. While $N$ does not directly affect the computational cost, long auto-correlation times due to a low acceptance rate could indirectly extend the runtime. Nevertheless, we find that this model shows only a weak dependence of the acceptance rate ($\sim 40$--$50\%$) on $N$ (Appendix \ref{app:QMC_details}, Figure~\ref{fig:acceptance}), making it well suited for DQMC simulations at intermediate to large values of $N$. Lastly, we are employing a higher-order Trotter decomposition to improve the systematic Trotter error \cite{WangAssaad21,Goth2022}.

In contrast to exact diagonalisation (ED) or DMRG calculations where one can directly access the energies of the lowest eigenstates, for QMC the critical data must be measured through correlation functions of (local) observables. On the fuzzy sphere, the simplest observables are the fermion-bilinear density operators. Here, the bilinears can be organised in different representations of the $\Sp(N)$ global symmetry, namely, the singlet $S$, the anti-symmetric traceless rank-2 tensor $A$ and symmetric rank-2 tensor $T$
\begin{subequations}
\begin{align}
    n_S(\br)&=\psi^\dagger_i\psi^i\\
    n_A^{[ij]}(\br)&=\psi^\dagger_k\Omega^{ki}\psi^j-\psi^\dagger_k\Omega^{kj}\psi^i-\tfrac{1}{N}\Omega^{ij}\psi^\dagger_k\psi^k \label{eq:def_den_a}\\
    n_T^{(ij)}(\br)&=\psi^\dagger_k\Omega^{ki}\psi^j+\psi^\dagger_k\Omega^{kj}\psi^i.
\end{align}
\end{subequations}
Under the particle-hole symmetry $\mathcal{P}$, $n_S$ and $n_A$ are odd while $n_T$ is even. The density operators can be expressed in terms of the angular components $n_{R,lm}$
\begin{align}
    n_{O,lm}&=\int\rd^2\br\,\bar{Y}_{lm}(\br)n_O(\br)\nonumber\\
    n_O(\br)&=\frac{1}{R^2}\sum_{lm}n_{O, lm}Y_{lm}(\br).
\end{align}
where $O=S,A,T$ is the $\Sp(N)$ representation. 

With quantum Monte Carlo, we can measure their real-space equal-time two-point functions
\begin{equation}
    \cC_O(\gamma_{12})=\langle (n_O(\br_1))^\dagger n_O(\br_2)\rangle
    \label{twopoint}
\end{equation}
where $\gamma_{12}$ is the angular distance between the two points. It can be decomposed into the linear combination of the correlators of the angular components $\langle n_{O,lm}^\dagger n_{O,lm}\rangle$. The details are given in Appendix~\ref{app:den}. We can also measure the imaginary-time-displaced correlator of a given angular component
\begin{equation}
    \cC_{O,l}(\tau)=\langle (n_{O,l0}(\tau))^\dagger n_{O,l0}(0)\rangle
\end{equation}
Here we have taken $m=0$ because all the choices of $m$ are equivalent due to the rotation symmetry. 

\section{Results}

\subsection{Phase Diagram} 

\begin{figure}
    \centering
    \includegraphics[width=0.45\textwidth]{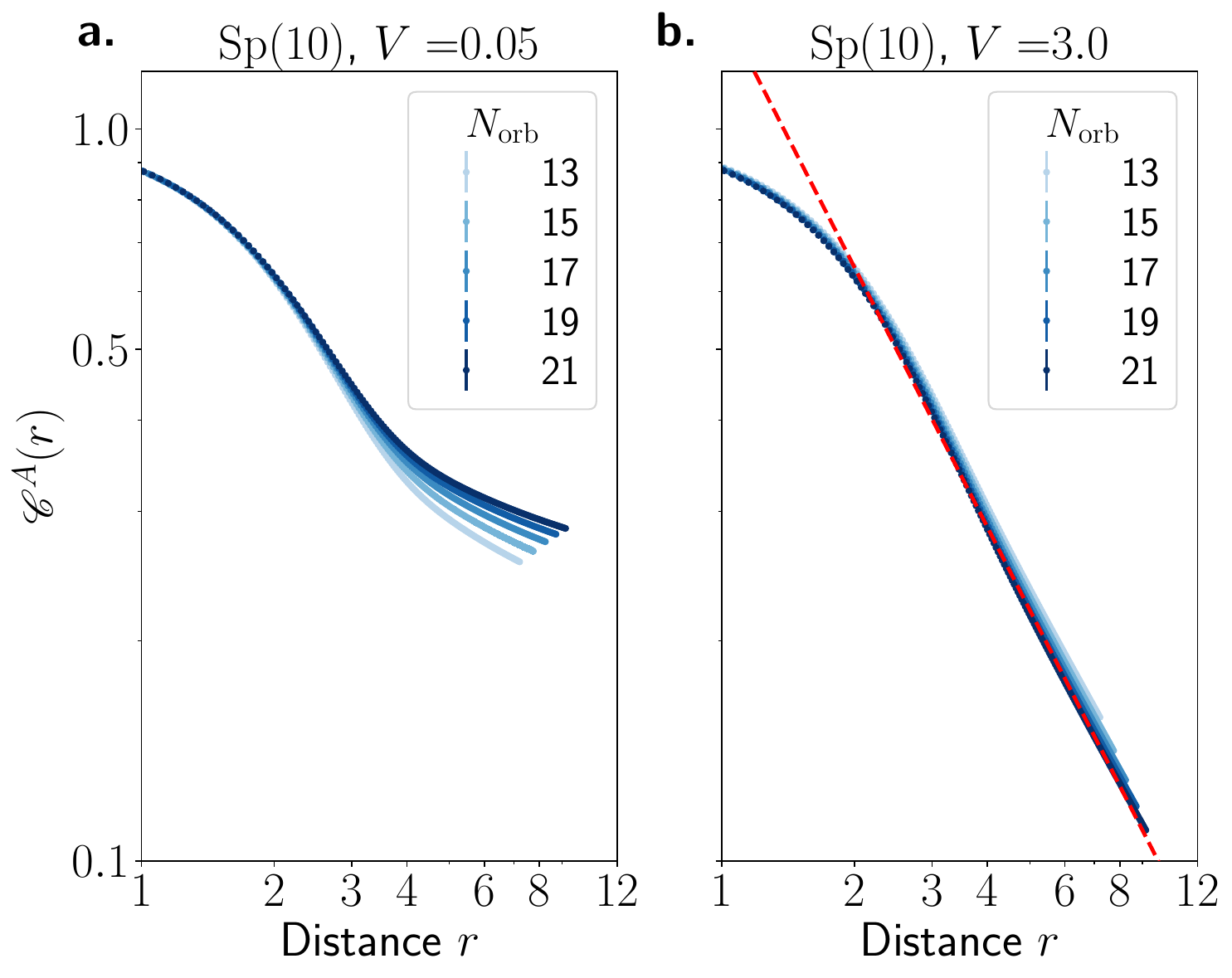}
    \caption{The equal-time correlation functions in the $\mathrm{Sp}(10)$ model at (a) $V=0.05$ in the spontaneous-symmetry-broken phase and (b) $V=3.0$ in the critical QCD phase. The linear distance is taken as $r=2R\sin(\gamma_{12}/2)$. The values of $C^a(r)$ are scaled so that $C^a(0)=1$ for each of the system sizes. The red dashed line in (b) marks the a power law fit to data for $r>2.5$.} 
    \label{conformalform10}
\end{figure}

\begin{figure*}
    \centering
    \includegraphics[width=.32\textwidth]{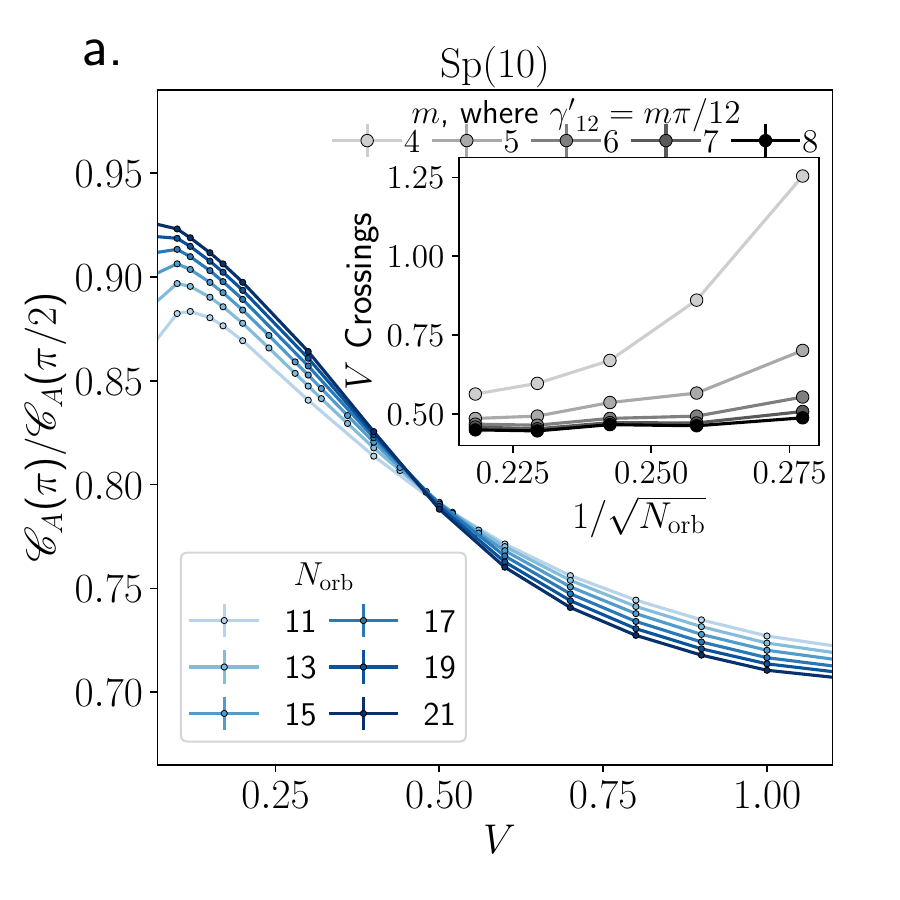}
    \includegraphics[width=.32\textwidth]{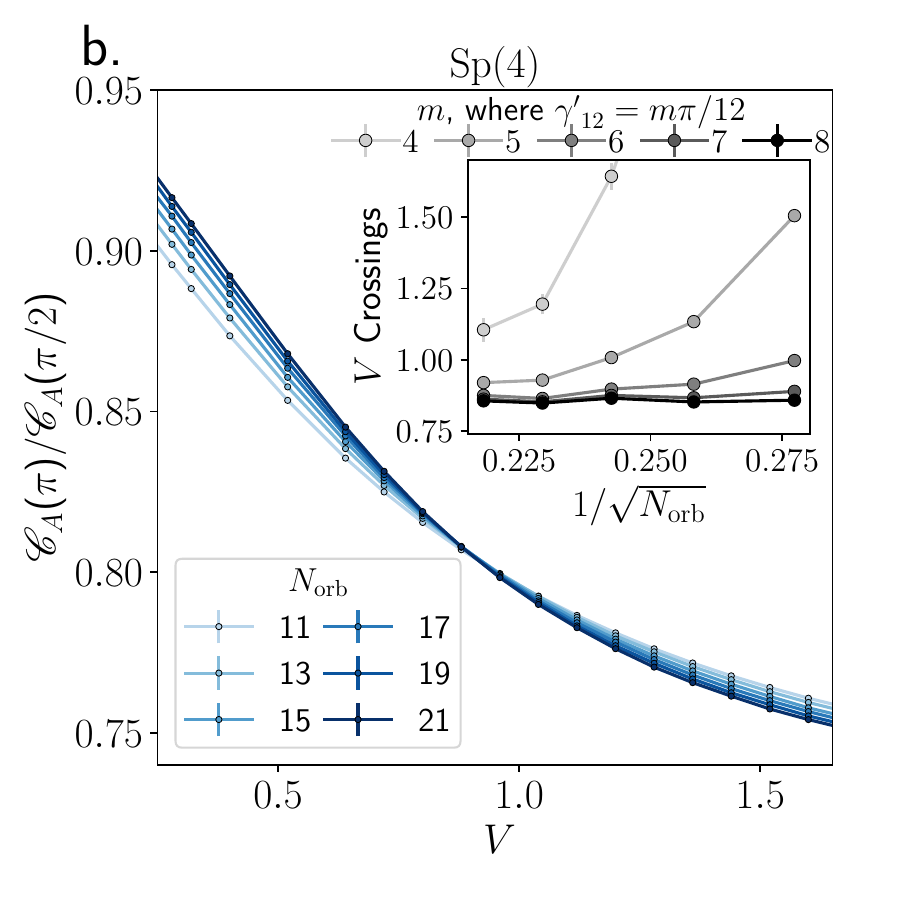}
    \includegraphics[width=.32\textwidth]{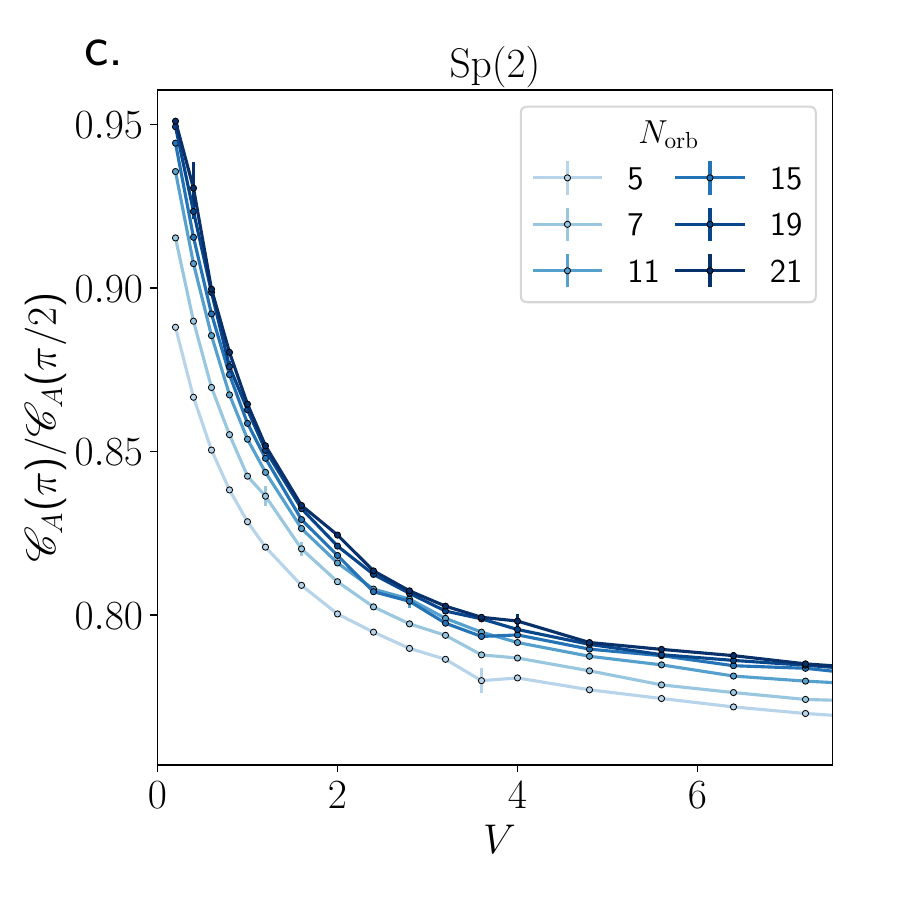}
    \caption{\textit{Main:} RG-invariant observables $\cC_A(\pi) / \cC_A(\pi/2)$ as a function of the coupling $V$ for $\mathrm{Sp}(10)$, $\mathrm{Sp}(4)$, and $\mathrm{Sp}(2)$ models. \textit{Insets of (b) and (c):} The $V$-value at which the RG-invariant observable $\cC_A(\pi) / \cC_A(\gamma'_{12}=m\pi/12)$ has a crossing as a function of $1/\sqrt{N_{\mathrm{orb}}}$ for pairs $(N_{\mathrm{orb}}-2,N_{\mathrm{orb}})$. } 
    \label{rginv4_10}
\end{figure*}

We first determine the phase diagram through the equal-time correlation functions and related RG-invariant dimensionless quantities, in parallel with the standard analysis used in lattice models. 

The order parameter that captures the symmetry-breaking pattern on the Grassmannian~\eqref{eq:grass} is the anti-symmetric density operator~\eqref{eq:def_den_a} $n^A_{[ij]}$. Taking the $\Sp(10)$ theory as an example, we calculate its two-point function~\eqref{twopoint} $\cC_A(\gamma_{12})$ and normalise it by its value at zero distance $\cC_A(\gamma_{12}=0)$~\footnote{The zero-distance correlator $\cC_A(0)$, which diverges in the CFT limit, is instead controlled by model-dependent short-range behaviour in the present model and thus saturates to a constant in the thermodynamic limit.} 
\begin{equation*}
    \frac{\cC_A(r)}{\cC_A(0)}=\frac{\langle n_{A,[ij]}(\br_1)n_A^{[ij]}(\br_2)\rangle}{\langle n_A(\br)^2\rangle}.
\end{equation*}
We plot it in Figure~\ref{conformalform10} as a function of the spatial distance $r=2R\sin(\gamma_{12}/2)$ for different system sizes, where $\gamma_{12}$ is the angular distance between the two points, and $R=\sqrt{N_\orb}$. We identify two likely distinct phases: At $V=0.05$, the correlation function tends to remain finite at long distance, indicating a spontaneous-symmetry-broken (SSB) phase; at $V=3$, the data at large distance converges to a power law, indicating a critical phase with scale-invariance. This phase diagram is consistent with the conformal window depicted in Figure~\ref{fig:ph_diag}a. The critical phase flows to the stable QCD fixed point.

To determine the critical point between the two phases, we examine the ratio of the correlator at two distinct distances 
\begin{equation*}
    \lambda=\cC_A(\gamma_{12})/\cC_A(\gamma_{12}').
\end{equation*}
This quantity is dimensionless and is RG-invariant at fixed points. Similar to the Binder cumulant~\cite{BINDER198879}, this ratio approaches distinct values in different phases and exhibits crossings for different system sizes at a critical point~\cite{Campostrini14}. We take $\gamma_{12}=\pi$ in the numerator as the anti-podal correlation. For both $\Sp(10)$ and $\Sp(4)$, we observe such an intersection marking a critical point $V_c$~(Figure~\ref{rginv4_10}a,b). At $V<V_c$, $\lambda$ increases with system size and approaches unity as expected for the symmetry broken phase; for $V>V_c$, $\lambda$ decreases with system size and converges to a different finite value. We extract the crossing point $V_c$ for different size pairs $(N_\orb-2,N_\orb)$ at multiple choices of $\gamma_{12}'$~(Figure~\ref{rginv4_10}a,b, insets).\footnote{The crossing points and their errors are estimated using a polynomial fit of the curves in their vicinity.} For all $\gamma_{12}'=m\pi/12$ ($m=4,5,6,7,8$), the crossing point converges to the same value --- $V_c\approx 0.5$ for $\Sp(10)$ model and $V_c\approx 0.8$ for the $\Sp(4)$ model --- with increasing system size. This result suggests a continuous phase transition with scale-invariance at the crossing point separating a SSB phase and a critical QCD phase (Figure~\ref{fig:ph_diag}a). For comparison, $\lambda$ increases monotonously with system size at all parameter ranges with no crossing point in the $\Sp(2)$ model~(Figure~\ref{rginv4_10}c), suggesting a symmetry-broken phase throughout the phase diagram. This is consistent with the pseudo-criticality for the $\SO(5)$ DQCP~\cite{Zhou2023SO5} (Figure~\ref{fig:ph_diag}c).

\begin{figure*}[tp]
    \centering
    \includegraphics[width=0.99\textwidth]{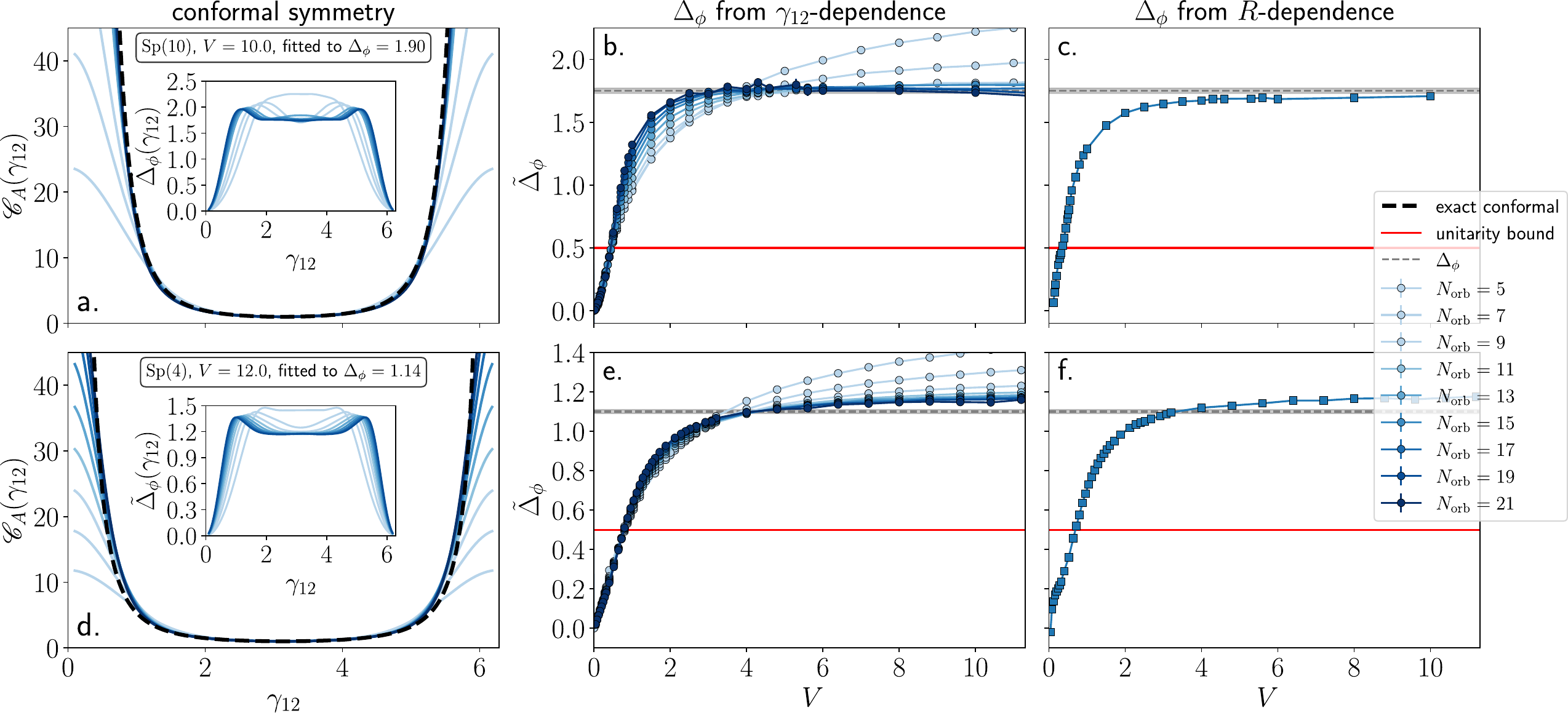}
    \caption{The conformal correlator $\cC_A(\gamma_{12})$ and the extracted scaling dimension $\Delta_\phi$ for (a--c) the $\Sp(10)$ model and (d--f) the $\Sp(4)$ model. (a,d) The real-space correlator~\eqref{eq:conf_corr} $\cC_A(\gamma_{12})$ of the density operator $n_A$ as a function of the angular distance $\gamma_{12}$, measured at (a) $V=10$ and (d) $V=12$. The dashed lines mark the best-fit conformal 2-pt function. \textit{Insets of (a,d).} The scaling dimension extracted through Eq.~\eqref{eq:dim_corr} as a function of $\gamma_{12}$. The dashed line marks the best-fit conformal 2-pt function. (b,e) The scaling dimension $\Delta_\phi$ as a function of $V$ extracted through $\gamma_{12}$-dependence~\eqref{fssscale}. (c,f) The scaling dimension $\Delta_\phi$ as a function of $V$ extracted through $R$-dependence~\eqref{eq:dim_corr}. The dashed grid-line marks and the grey shade marks the finite-size scaling result of $\Delta_\phi$ and its error-bar (see Appendix~\ref{app:fit_dim}). The red grid-line marks the unitarity bound $\Delta=1/2$.} 
    \label{conformalform4}
\end{figure*}

\subsection{Conformal Correlator and Scaling Dimensions}

Given the evidence for a scale-invariant critical point, we present evidence for the conformal symmetry and extract the conformal data in the critical QCD phase $V>V_c$ from the same equal-time two-point function for the $\Sp(10)$ and $\Sp(4)$ models.

At a conformal fixed point, any gapless local observables $n_O(\br)$ on the fuzzy sphere can be written as the linear combination of CFT operators in the same representation of global symmetry and parity. To the lowest order, $n_A$ and $n_T$ are used to express the fermion bilinear operator~\eqref{eq:op_antisym} in the CFT and the temporal component of the conserved symmetry current. 
\begin{align}
    n_A(\br)&=\text{const.}\times \phi(\br)+\dots\nonumber\\
    n_T(\br)&=\text{const.}\times J^\tau(\br)+\dots
    \label{eq:den_cft_op}
\end{align}
To the leading order, the two-point correlator $\cC_A(\gamma_{12})$ should behave like a power-law~\cite{Han2023Jun,Cardy1996Scaling}
\begin{eqnarray}
    \cC_A(\gamma_{12})&=&\langle n_A(\br_1)n_A(\br_2)\rangle\nonumber\\
    &\overset{\text{CFT}}{\mathop{=}}&\text{const.}\times\langle\phi(\hat{\mathbf{n}}_1)\phi(\hat{\mathbf{n}}_2)\rangle_{S^2\times\mathbb{R} }\nonumber\\
    &=&\text{const.}\times R^{-2\Delta_\phi}\left(2\sin\frac{\gamma_{12}}{2}\right)^{-2\Delta_\phi}.
    \label{eq:conf_corr}
\end{eqnarray}
where $\hat{\mathbf{n}}_i={\mathbf{r}}_i/r_i$ is the unit vector. We note that the $R$-dependence is a consequence of scale invariance, while the $\gamma_{12}$-dependence is a consequence of conformal invariance. We measure this two-point function through the equal-time correlator in the QMC. To compare with the conformal  form, we cancel the constant factor by normalising $\cC_A(\gamma_{12})$ by its value at $\gamma_{12}=\pi$, and compare $\cC_A(\gamma_{12})/\cC_A(\pi)$ with the conformal correlator $\left(\sin\frac{\gamma_{12}}{2}\right)^{-2\Delta_\phi}$. With increasing system size, the measured correlator on the fuzzy sphere converges towards the conformal correlator (Figure~\ref{conformalform4}a,d). 

As an important piece of conformal data, the scaling dimension $\Delta_\phi$ can be extracted from the conformal correlator. Starting from Eq.~\eqref{eq:conf_corr}, we employ two methods:
\begin{enumerate}
    \item Fix the size $R$ and examine the dependence on the angular distance $\gamma_{12}$. The $\Delta_\phi$ can be measured through the logarithmic derivative of the correlator
    \begin{align}
        \tilde{\Delta}_\phi(\gamma_{12},V,R)&=-\frac{1}{2}\frac{\partial\log \cC_A(\gamma_{12})}{\partial\log r_{12}}\nonumber\\
        &=-\tan \frac{\gamma_{12}}{2}\frac{\partial\log \cC_A(\gamma_{12})}{\partial\gamma_{12}}
        \label{eq:dim_corr}
    \end{align}
    where $r_{12}=2R\sin(\gamma_{12}/2)$ is the conformal distance. We emphasize that this method of extracting $\Delta_\phi$ utilizes the power of conformal symmetry, such that the traditional finite size scaling with respect to system size is not needed~\cite{Cardy1996Scaling}.
    \item Fix the angular distance $\gamma_{12}$ and examine the dependence on the size $R$. At a certain $\gamma_{12}$,
    \begin{equation}
        \cC_A(\gamma_{12}) \sim R^{-2\Delta_\phi},
        \label{fssscale}
    \end{equation}
    so a power-law fit of $\cC_A$ as a function of $R=N_\orb^{1/2}$ directly yields $\Delta_\phi$.
\end{enumerate}

The $\Dlt_\phi$ measured through $\gamma_{12}$-dependence is in general a function of angle $\gamma_{12}$, parameter $V$ and system size $R$. As a quantitative check that the measured correlator exhibits conformal symmetry, we first show that its value is approximately independent of $\gamma_{12}$ over a wide range of distances, with deviations appearing only in the short-distance region $\gamma_{12}\ll 1$ dominated by the regulator (Figure~\ref{conformalform4}a,d, insets). We find that in extracting  $\Delta_\phi$ from Eq.~\eqref{eq:dim_corr}, $\gamma_{12}$ slightly smaller than $\pi$ empirically suffers from the small finite size effect and QMC statistical error. Hereafter, we fix $\gamma_{12} = 7\pi / 9$. We examine the dependence of $\tilde{\Delta}_\phi$ on $V$ (Figure~\ref{conformalform4}b,e). For different values of $V$ within the critical QCD phase, the extracted scaling dimension scales towards a common value in the thermodynamic limit. Performing a finite-size scaling analysis across multiple $V$ values, we obtain $\Delta_\phi=1.10(1)$ for $\Sp(4)$ and $1.75(2)$ for $\Sp(10)$. Further detail of the finite-size scaling is provided in Appendix~C. As an additional consistency check, in the symmetry-broken phase, this analysis yields $\tilde{\Delta}_\phi<1/2$ violating the unitarity bound as expected. From the alternative method based on $R$-dependence, the fitted scaling dimension $\Delta_\phi$ is consistent with those extracted from $\gamma_{12}$-dependence (Figure~\ref{conformalform4}c,f).

\begin{figure}[ht]
    \centering
    \hfill\begin{minipage}[c]{0.3\textwidth}
        \includegraphics[width=\linewidth]{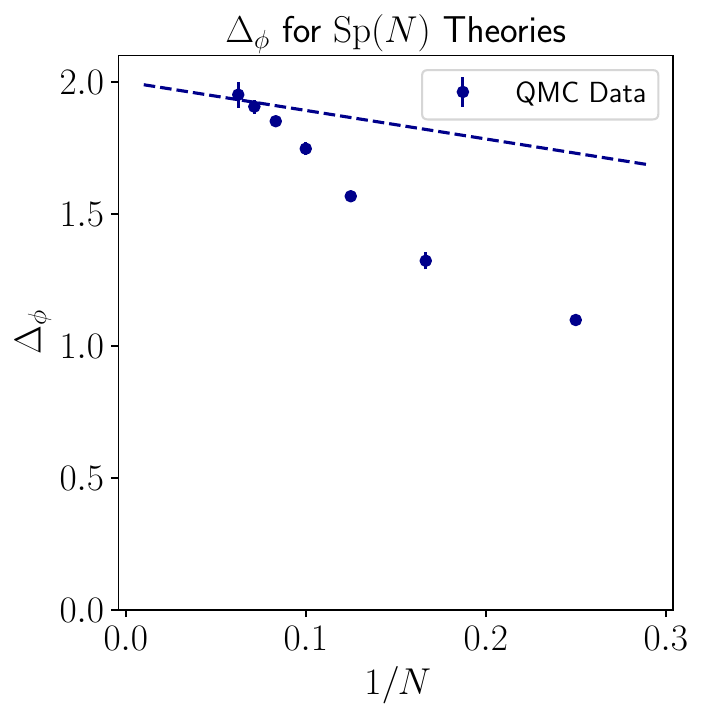}
    \end{minipage}\hfill
    \begin{minipage}[c]{0.18\textwidth}
        \centering
        \setlength{\tabcolsep}{6pt}
        \renewcommand{\arraystretch}{1.2}
        \begin{tabular}{rl}
            \hline\hline
            $N$ & $\Delta_\phi$ \\
            \hline
            $4 $ & $1.10(1)$ \\
            $6 $ & $1.32(3)$ \\
            $8 $ & $1.57(1) $ \\
            $10$ & $1.75(2)$ \\
            $12$ & $1.85(2)$ \\
            $14$ & $1.90(2)$ \\
            $16$ & $1.95(5) $ \\
            \hline\hline
        \end{tabular}
    \end{minipage}\hfill{}
    \caption{The scaling dimension $\Delta_\phi$ as a function of $N$ calculated from the conformal correlator. The dashed line is the large-$N$ expansion result~\cite{Xu2008LargeN}. }
    \label{largenplot}
\end{figure}

As discussed above, the computational cost in the QMC simulations depends only weakly on the number $N$ of flavours. Thanks to this advantage, we are able to access the model at large $N$ up to $N=16$. For each $N$, we extract the scaling dimension $\Delta_\phi$ through Eq.~\eqref{eq:dim_corr}, for details see Appendix \ref{app:fit_dim}. With increasing $N$, the results approach the large-$N$ expansion (Figure~\ref{largenplot}).

\begin{figure}[t]
    \centering
    \includegraphics[width=.48\textwidth]{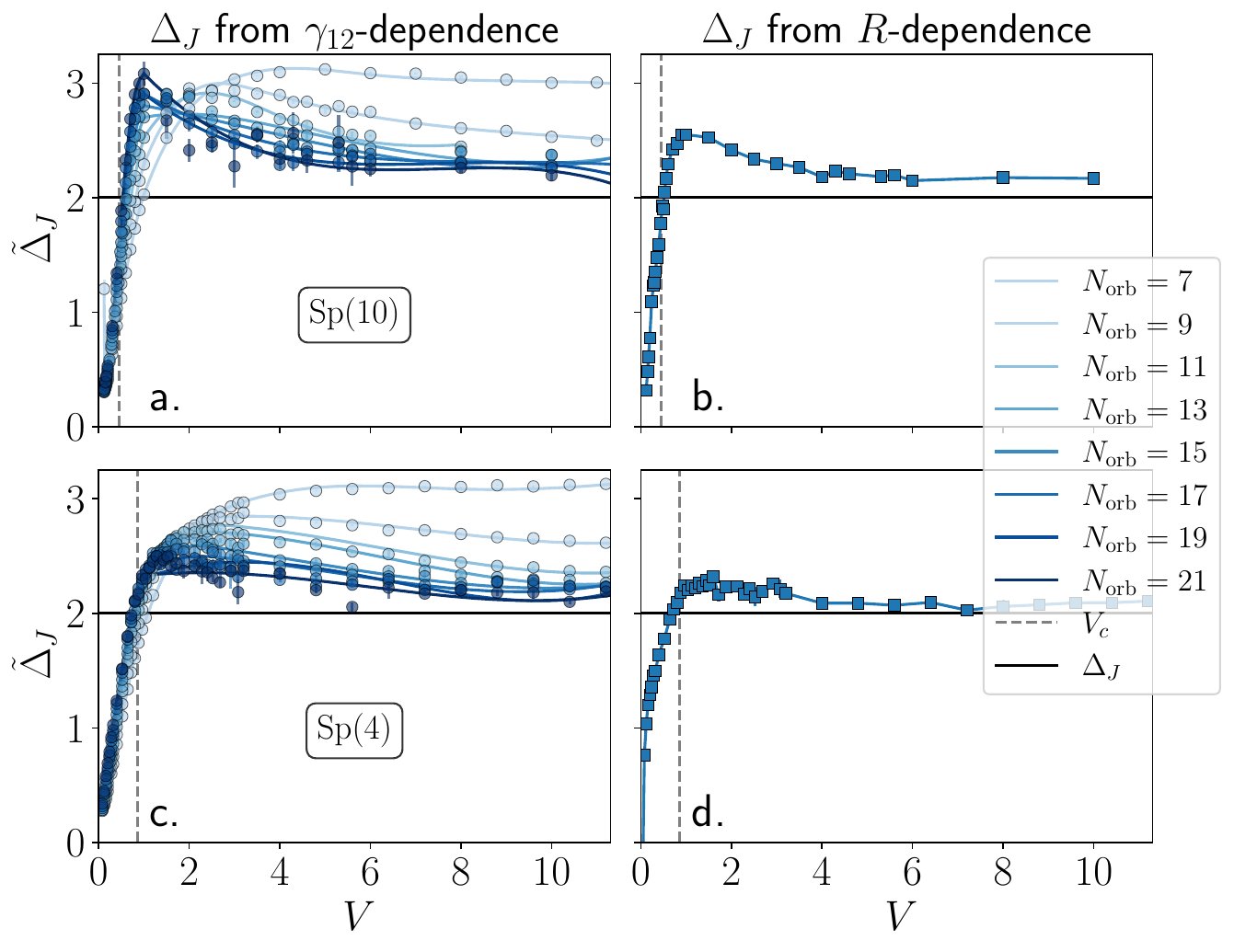}
    \caption{The scaling dimension $\Delta_\phi$ as a function of $V$ calculated through conformal 2-pt correlator $\cC_T$ in the (a,b) $\Sp(10)$ and (c,d) $\Sp(4)$ models. Panels (a) and (c) are extracted through $\gamma_{12}$-dependence; (b) and (d) are extracted through $R$-dependence. The black line marks the expected value $\Delta_J=2$. The dashed grey line marks the critical point extracted through crossing in Figure~\ref{rginv4_10}.} 
    \label{scalingcollapses3}
\end{figure}

We have also checked the correlator in the symmetric representation $T$, which corresponds to the two-point function of the conserved symmetry current $J^\mu$ with scaling dimension $\Delta_J=2$
\begin{equation}
    \cC_T(\gamma_{12})
    \mathop{\sim}^\text{CFT} \langle J^\tau(\hat{\mathrm{n}}_1)J^\tau(\hat{\mathrm{n}}_2)\rangle_{S^2\times\mathbb{R}}
    \sim \left(2R\sin\frac{\gamma_{12}}{2}\right)^{-4}.
\end{equation}
where $\sim$ denotes the omission of a constant factor. We can extract the scaling dimension $\Delta_J$ in a similar fashion as Eqs.~\eqref{eq:dim_corr} and \eqref{fssscale}. In the critical QCD phase, the scaling dimension approaches $\Delta_J=2$ in the thermodynamic limit as expected (Figure~\ref{scalingcollapses3}). When crossing the critical point from the SSB phase into the critical QCD phase, $\Delta_J$ first increases from $\Delta_J<2$ to $\Delta_J>2$ and then gradually relaxes back to $2$ deeper in the conformal phase. Notably, the coupling at which $\Delta_J$ crosses $2$ lies close to the critical point extracted from the RG-invariant analysis in Figure~\ref{rginv4_10}, which is suggestive of a conformal transition.

\begin{figure*}
    \centering
    \includegraphics[width=.45\textwidth]{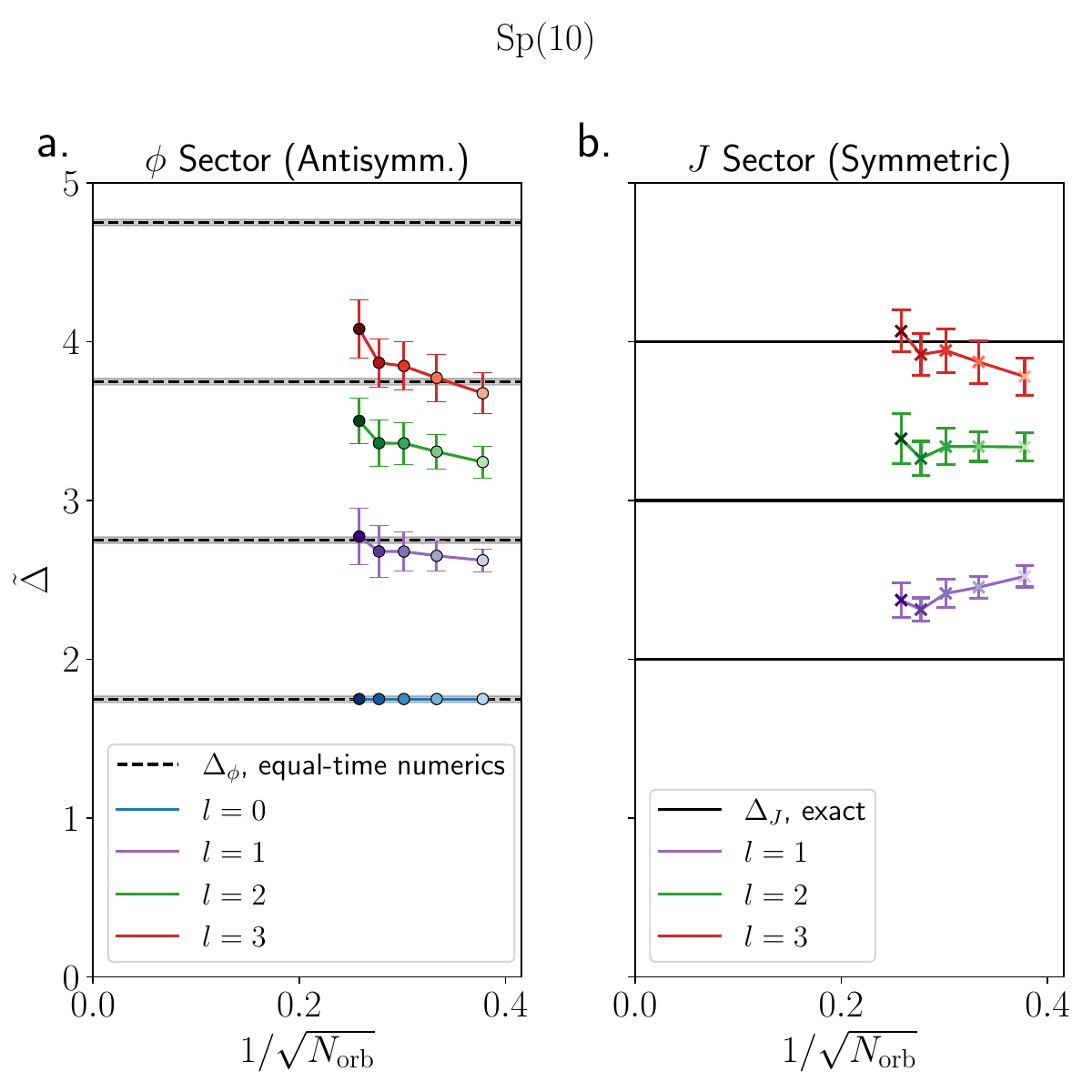}
    \includegraphics[width=.45\textwidth]{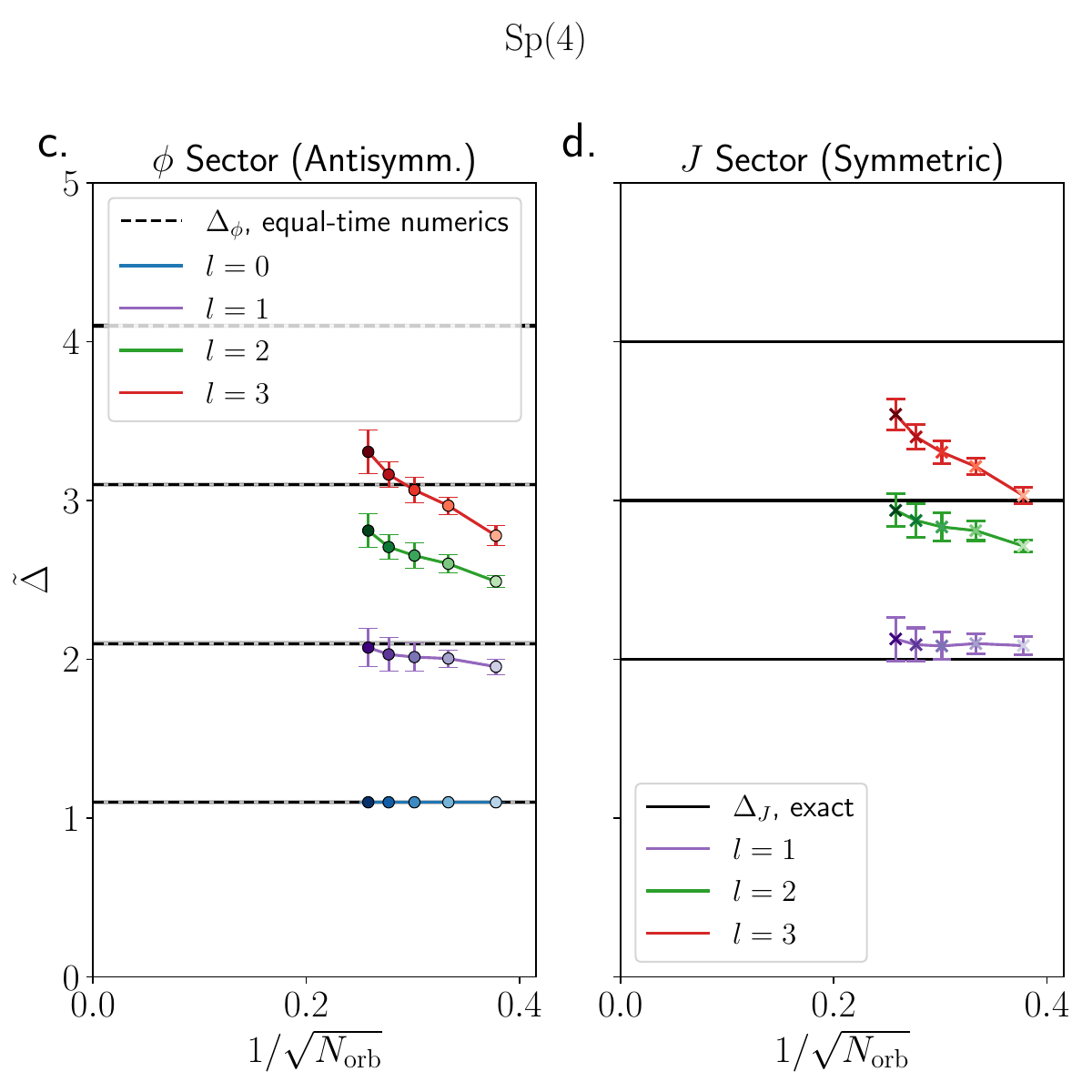}
    
    \caption{The operator spectrum by state-operator correspondence for the lightest operators in each symmetry sector as a function at different system size $N_\orb$. The data are extracted through the time-displaced correlator for (a,b) the $\Sp(10)$ model at $V=5.0$ and (c,d) the $\Sp(4)$ model at $V=6.0$.  Panels (a) and (c) plots the multiplet $\partial^l\phi$ of $\phi$ in the $A$ sector; (b) and (d) plots the multiplet $\partial^{l-1}J$ of $J$ in the $T$ sector.}
    \label{conformalform10_2}
\end{figure*}

\subsection{State-Operator Correspondence}

On the fuzzy sphere, we can make use of the state-operator correspondence to further verify conformal symmetry and extract scaling dimensions. Each eigenstate of the Hamiltonian at the conformal fixed point has a one-to-one correspondence to a CFT local operator. The state and the operator has corresponding Lorentz spin, representation under global symmetry, and the excited energy is proportional to the scaling dimension 
\begin{equation}
    E_\Phi-E_0=\frac{v}{R}\Delta_\Phi
    \label{eq:st_op_corr}
\end{equation}
where $v$ is a model-dependent light speed. For quantum Monte Carlo, the excited states can be accessed through the time-displaced correlation functions of the angular components of the density operator $n_{O,l0}$ in certain representation $O=S,A,T$ of the $\Sp(N)$ symmetry. At zero-temperature, it can be expressed in terms of the spectral decomposition 
\begin{align}
    \cC_{O,l}(\tau)&=\langle (n_{O,l0}(\tau))^\dagger n_{O,l0}(0)\rangle\nonumber\\
    &=\sum|\langle\Phi_{O,l}|n_{O,l0}|0\rangle|^2e^{-(E_\Phi-E_0)\tau}
    \label{spectral}
\end{align}
where the sum is taken over the excited states $|\Phi_{O,l}\rangle$ in the $O$ representation with Lorentz spin $l$. To the leading order at large $\tau$, the rate of the exponential decay yields the scaling dimension $\Delta_{O,l}$ of the the lowest scaling operator in the corresponding sector
\begin{equation}
    \cC_{O,l}(\tau)=\text{const.}\times e^{-(v/R)\tau\Delta_{O,l}}+\text{subleading}\,.
\end{equation}
The subleading terms encode other operators in the same representation with the same Lorentz spin, \textit{e.\,g.}, certain descendants like $\partial_\mu \partial^\mu O_l$ which have a higher scaling dimension and therefore decay faster. In the anti-symmetric representation $A$, the lowest states are the primary $\phi$ and its descendants $\partial^{\mu_1}\cdots\partial^{\mu_l}\phi$; in the symmetric reprensentation $T$, the lowest states are the conserved symmetry current $J^\mu$ and its descendants $\partial^{\mu_1}\cdots\partial^{\mu_{l-1}}J^{\mu_l}$. Conformal symmetry thus predicts
\begin{align*}
    \Delta_{A,l}&=\Delta_\phi+l,&\Delta_{T,l}&=1+l.
\end{align*}

We perform the analysis for the $\Sp(10)$ and $\Sp(4)$ models and extract $\Delta_{A,l}$ for $l=0,1,2,3$ and $\Delta_{T,l}$ for $l=1,2,3$ (Figure~\ref{conformalform10_2}) in the conformal phase. The model-dependent velocity $v$ is fixed by calibrating the lowest antisymmetric gap $\Delta_{A,0}=\Delta_\phi$, using the value obtained from equal-time conformal correlators. With increasing system size, the extracted scaling dimensions in both sectors for both models approach the conformal predictions, providing further evidence for emergent conformal symmetry.

\section{Discussion}

In this paper, we have identified a family of CFTs on the fuzzy sphere which putatively corresponds to the $\SU(2)$ quantum chromodynamics (QCD) with $N$ flavours of fermions in three space-time dimensions. It is a generalisation of the $\SO(5)$ deconfined criticality at $N=2$. On the fuzzy sphere, we construct a model with $N_f=2N$ flavours of fermions and $\Sp(N)$ global symmetry. We show that the model matches the symmetry and anomaly with a $\Sp(N)$-symmetric non-linear sigma model (NLSM) with a level-1 Wess-Zumino-Witten (WZW) topological term. Extending the NLSM to the strongly-coupled region, its phase diagram depends on $N$. For $N > N_c$, defining the conformal window, the theory exhibits a symmetry-broken phase and a critical QCD phase separated by a continuous phase transition. For $N < N_c$, the phase diagram contains only a symmetry-broken phase and may display pseudo-critical behaviour in certain parameter regions, as putatively expected for the DQCP. 

This fuzzy-sphere model is free of the sign problem and therefore accessible with large-scale quantum Monte Carlo simulations. For $N\geq 4$, we find evidence for the phase diagram containing a symmetry-broken phase and a critical QCD phase separated by a continuous phase transition. By computing real-space equal-time correlators as well as time-displaced correlators that probe excited states, we identify signatures of conformal symmetry in the critical phase, including conformal two-point functions and integer-spaced conformal multiplets. From the conformal correlator, we extract the scaling dimension $\Delta_\phi$ of the leading operator in the rank-2 traceless anti-symmetric tensor representation, which serves as the order parameter for the phase transition. Notably, $N$ enters the QMC formulation as a parameter instead of the number of fermions simulated, allowing us to reach up to $N=16$. Our results on the fuzzy sphere in the large-$N$ are consistent with the perturbative large-$N$ calculation.

This agreement between fuzzy sphere and large-$N$ expansion provides additional evidence that the conformal field theory observed on the fuzzy sphere indeed corresponds to $\SU(2)$ QCD$_3$. In earlier fuzzy-sphere studies of the $\SO(5)$ DQCP~\cite{Zhou2023SO5} and $\Sp(N)$ CFTs~\cite{Zhou2024Oct}, the matching with candidate Lagrangian descriptions was limited to the global symmetries and anomalies. Due to the strongly coupled nature of these theories, a more quantitative comparison was not previously available. Our work represents the first extension of the fuzzy-sphere approach into the large-$N$ region, where the CFT becomes weakly coupled and the perturbative large-$N$ expansion is quantitatively reliable. In this region, we are able to demonstrate agreement between fuzzy-sphere results and large-$N$ predictions, providing evidence that the family of $\Sp(N)$-symmetric fuzzy-sphere models realises the candidate Lagrangian description of $\SU(2)$ QCD$_3$. Meanwhile, in the large-$N$ limit, the fuzzy-sphere model may exhibit semi-classical behaviour, and it may be interesting to investigate it perturbatively using approaches like saddle-point expansion.

Returning to the strongly coupled regionf at finite $N$, we have presented evidence that $\SU(2)$ QCD$_3$ is conformal for flavor number $N \ge 4$. Combined with previous results indicating that the DQCP at $N=2$ is not genuinely critical, our findings tentatively suggest that the boundary of the conformal window lies in the range $2<N_c<4$. A sharper diagnostic of the conformal window is provided by the scaling dimension of the singlet operator $S_+$, which governs the RG flow in the phase diagram of the extended NLSM (Figure~\ref{fig:ph_diag}). For $N > N_c$, $S_+$ is irrelevant at the stable QCD fixed point and with $\Delta_{S_+} > 3$; for $N < N_c$, $S_+$ becomes relevant with $\Delta_{S_+} < 3$~\cite{Zhou2023SO5}; and at the critical value $N = N_c$, $S_+$ is exactly marginal with $\Delta_{S_+} = 3$. Determining the scaling dimension $\Delta_{S_+}$ would therefore provide a decisive criterion for the conformal-window boundary $N_c$. However, the operator $S_+$ can be realised only by a four-fermion term rather than a two-fermion density operator, and its measurement requires more sophisticated QMC techniques. Since $N$ enters the QMC formulation as a continuous parameter, it is in principle possible to analytically continue $N$ to non-even-integer values and determine $\Delta_{S_+}(N)$ as a smooth function of $N$, opening the possibility of directly observing $\Delta_{S_+}(N_c)=3$.

While this work focuses on the conformal window of $\SU(2)$ QCD$_3$, analogous conformal-window problems arise more broadly in three-dimensional critical gauge theories involving multiple flavors of fermions or scalars coupled to gauge fields, such as $\SU(k)$, $\rU(k)$, or $\Sp(k)$. Some of these theories may be realisable on the fuzzy sphere through appropriate NLSM-WZW constructions, while others, such as $\rU(k)$ gauge theories, have additional challenges realising their $\rU(1)$ flux symmetry. Further discussion is provided in Appendix~\ref{app:wzw}. An especially important open problem is the conformal window of QED$_3$ with $N$ flavors of Dirac fermions, which is directly tied to the nature of the Dirac spin liquid at $N=4$.

\FloatBarrier

\section*{Acknowledgments}

We would like to thank Francesco Parisen Toldin, Karl Jansen, and Anders Sandvik for fruitful discussions. Research at Perimeter Institute is supported in part by the Government of Canada through the Department of Innovation, Science and Industry Canada and by the Province of Ontario through the Ministry of Colleges and Universities. Z.~Z.~acknowledges support from the Natural Sciences and Engineering Research Council of Canada (NSERC) through Discovery Grants. 
J.~H. acknowledges financial support by the Deutsche Forschungsgemeinschaft (DFG, German Research Foundation) through the W\"urzburg-Dresden Cluster of Excellence ctd.qmat --- Complexity, Topology and Dynamics in Quantum Matter (EXC 2147, project-id 390858490)
and via the project A07 of the Collaborative Research Center SFB 1143 (Project No.~247310070).
The auxillary field QMC simulations were carried out with the ALF 
package~\cite{alf-v2.4,alf-v2.0} 
available at \url{https://alf.physik.uni-wuerzburg.de}.
The WIGXJPF library was used for the 3$j$-symbols~\cite{Johansson_2016}.

\vspace*{0.2cm}

\appendix

\section{The Angular Components of the Density Operators}
\label{app:den}

The density operator is local fermion bilinear
\begin{equation}
    n_M(\hat{\mathbf{n}})=\psi_i^\dagger(\hat{\mathbf{n}})M^i{}_j\psi^j(\hat{\mathbf{n}}).
    \label{eq:den_def}
\end{equation}
Here, the matrix insertion $M$ puts the density operators in a certain representation of the flavour symmetry. 

Like the fermion operator, the density operator can also be expressed in the orbital space
\begin{equation}
    n_M(\hat{\mathbf{n}})=\frac{1}{R^2}\sum_{lm}Y_{lm}(\hat{\mathbf{n}})n_{M,lm}.
    \label{eq:den_decomp}
\end{equation}
Conversely,
\begin{widetext}
\begin{align}
    n_{M,lm}&=R^2\int\rd^2\hat{\mathbf{n}}\,\bar{Y}_{lm}(\hat{\mathbf{n}})n_M(\hat{\mathbf{n}})\nonumber\\
    &=\int\rd^2\hat{\mathbf{n}}\,\bar{Y}_{lm}(\hat{\mathbf{n}})\left(\sum_{m_1}\bar{Y}^{(s)}_{sm_1}(\hat{\mathbf{n}})c^\dagger_{m_1i}\right)M^i{}_j\left(\sum_{m_2}Y^{(s)}_{sm_2}(\hat{\mathbf{n}})c_{m_1}^j\right)\nonumber\\
    &=\sum_{m_1m_2}c^\dagger_{m_1i}M^i{}_jc_{m_2}^j\Lambda^{(n)}{}_{m_1m_2}^{lm}\\
    \Lambda^{(n)}{}_{m_1m_2}^{lm}&=\int\rd^2\hat{\mathbf{n}}\,\bar{Y}_{lm}(\hat{\mathbf{n}})\bar{Y}^{(s)}_{sm_1}(\hat{\mathbf{n}})Y^{(s)}_{sm_2}(\hat{\mathbf{n}})\nonumber\\
    &=\delta_{m+m_1,m_2}(-1)^{s+m_2}(2s+1)\sqrt{\frac{2l+1}{4\pi}}\begin{pmatrix}s&l&s\\m_1&m&-m_2\end{pmatrix}\begin{pmatrix}s&l&s\\s&0&-s\end{pmatrix}.
    \label{eq:den_mod}
\end{align}
\end{widetext}
where $\left(\begin{smallmatrix}\bullet&\bullet&\bullet\\\bullet&\bullet&\bullet\end{smallmatrix}\right)$ is the $3j$-symbol. In this way, we have fully expressed the density operator in terms of the operators in the orbital space $c^{(\dagger)}_{mf}$.

Similarly, we can decompose the pairing operators in terms of the monopole spherical harmonics
\begin{align}
    \Delta(\br)&=\frac{1}{R^2}\sum_{lm}Y_{lm}^{(2s)}\Delta_{lm}\nonumber\\
    \Delta_{2s,m}&=\sum_{m_1m_2}c_{m_1}^i\Omega_{ij}c_{m_2}^j\Lambda^{(\Delta)}{}_{m_1m_2}^{2s,m}\nonumber \\
    \Lambda^{(\Delta)}{}_{m_1m_2}^{2s,m}&=\delta_{m_1+m_2,m}(-1)^m\frac{2s+1}{\sqrt{4\pi}}\left(\begin{array}{ccc}s& s& 2s \\ m_1 & m_2 & -m \end{array}\right).
\end{align}
Here the only non-vanishing component is $l=2s$. From these components we can build the Hamiltonian Eq.~\eqref{eq:hamiltonian}.

We also write the correlation function of two points on the unit sphere in terms of the angular components
\begin{align}
    &\langle 0|n_M(\br_1)n_M(\br_2)|0\rangle\nonumber\\
    &=\sum_{lm_1m_2}\frac{1}{R^4}\langle n_{M,lm_1}n_{M,lm_2}\rangle Y_{lm_1}(\br_1)Y_{lm_2}(\br_2)\nonumber\\
    &=\sum_l\frac{2l+1}{R^2}P_l(\cos\theta_{12})\langle n_{M,l0}n_{M,l0}\rangle
\end{align}

\section{The Derivation of the WZW level}
\label{app:wzw}
\allowdisplaybreaks

\newcommand{\cD}{\mathcal{D}}
\newcommand{\cI}{\mathcal{I}}
\newcommand{\Qh}{\hat{Q}}
\newcommand{\bQh}{\hat{\mathbf{Q}}}
\newcommand{\bth}{\hat{\mathbf{t}}}
\newcommand{\Sgh}{\hat{\Sigma}}
\newcommand{\bG}{\mathbf{G}}
\newcommand{\Ah}{\hat{A}}
\newcommand{\bAh}{\hat{\mathbf{A}}}
\newcommand{\cb}{\bar{c}}
\newcommand{\bt}{\mathbf{t}}
\newcommand{\rp}{\mathrm{p}}
\newcommand{\rmq}{\mathrm{q}}
\newcommand{\phb}{\bar{\phi}}
\newcommand{\Skyr}{\textrm{Skyr}}

\subsection{General Principle}

We start from $N_f$ flavours of free fermions on the lowest Landau level on a large torus, the action reads 
\begin{align}
    Z&=\int\cD\psi\,\cD\psb\,e^{-S_\psi}\nonumber\\
    S_\psi[\psi,\psb]&=\int\rd^2\br\,\rd\tau\, \bar\psi_a(\br,\tau)\frac{\partial}{\partial\tau}\psi^a(\br,\tau). 
\end{align}
The fermion field are projected to the lowest Landau level
\begin{equation}
    \psi^a(\br,\tau)=\frac{1}{\sqrt{L_y}}\sum_{p_y}e^{ip_yy}\phi_{p_y}(x)c^a_{p_y}(\tau)
\end{equation}
The orbitals are labelled by the momentum along $y$ direction $p_y=2\pi m/L_y$ where $m=1,\dots,N_\phi=V/2\pi=L_xL_y/2\pi$, and the wavefunction in the limit of large $V$ is 
\begin{equation}
    \phi_{p_y}(x)=\frac{1}{\pi^{1/4}}\exp\left[-\half(x-p_y)^2\right]
\end{equation}

We then couple the free fermions to a matrix field $\bQh(\br,\tau)$ that lives on a Grassmannian. 
\begin{align}
    Z&=\int\cD\psi\,\cD\psb\,\cD\bQh\,e^{-S_\psi-S_Q}\nonumber\\
    S_Q[\psi,\psb,\bQh]&=-\lambda\int\rd^2\br\,\rd\tau\,\psb_a\Qh^a{}_b \psi^b
\end{align}
The standard parametrisation of the Grassmannian is 
\begin{gather}
    \bQ=\bg\Sigma\bg^{-1}\in\frac{\rG(N)}{\rG(N)\times\rG(N-M)}\\
    \bg\in\rG(N),\quad\Sigma=\begin{pmatrix}
        \BI_M&0\\0&-\BI_{N-M}
    \end{pmatrix}\nonumber\\
    \rG(N)=\Sp(N/2),\rU(N),\rO(N)\nonumber
\end{gather}
In vicinity of $\Sigma$, $\bQ$ can be expanded in terms of the generators $\bt_i$ of $\mathfrak{g}(N)$.
\begin{align}
    \bg&=1+\epsilon^i\bt_i\nonumber\\
    \bQ&=(1+\epsilon^i\bt_i)\Sigma(1-\epsilon^i\bt_i)\nonumber\\
    &=\Sigma+\epsilon^i\bA_i,\quad\bA_i=[\bt_i,\Sigma].
    \label{eq:lie_gen}
\end{align}

On the other hand, we consider a general set-up where the fermions on the LLL do not have to be $N$-flavour in the fundamental representation. \textit{E.~g.}, $N_f$ can be a multiplication of $N$, and the $N_f\times N_f$ matrix field $\bQh$ depends on the physical set-up of the fermions. The hat  denotes that $\bQh$ may be different from the standard parametrisation. We list several examples.
\begin{itemize}
    \item For $N_f=N$ flavours of fundamental fermions, $\bQh=\bQ$;
    \item for $N_f=2N$ flavours of fermions containing two copies of fundamental, $\bQh=\bQ\oplus\bQ$;
    \item for $N_f=2N$ flavours of fermions containing $N$ fundamental and $N$ anti-fundamental, $\bQh=\bQ\oplus\bQ^\dagger$.
\end{itemize}
Physically, this coupling captures certain four fermion interaction. Expanding this interaction in the momentum space
\begin{widetext}
\begin{align}
    S_Q[c,\cb,\bQh]&=-\lambda\int\rd\tau\sum_{p_y,\bq}F(p_y,\bq)\cb_{a,p_y+q_y/2}(\tau)\Qh^a{}_b(\bq,\tau)c^b_{p_y-q_y/2}(\tau)\nonumber\\
    F(p_y,\bq)&=\int\rd x\,e^{iq_xx}\phb_{p_y+q_y/2}(x)\phi_{p_y-q_y/2}(x)=e^{-q^2/4}e^{-iq_xp_y}
\end{align}
\end{widetext}
where we have used the Fourier transformation 
\begin{align*}
    c_{p_y}(\tau)&=\int\frac{\rd\omega}{2\pi}e^{i\omega\tau}c_{p_y}(\omega)\\
    \Qh^a{}_b(\br,\tau)&=\frac{1}{V}\sum_{\bq}\Qh^a{}_b(\bq,\tau)e^{i\bq\cdot\br}\nonumber\\
    &=\int\frac{\rd\nu}{2\pi}\frac{1}{V}\sum_{\bq}\Qh^a{}_b(\bq,\nu)e^{i\bq\cdot\br+i\nu\tau}
\end{align*}

The field $\bQh(\br,\tau)$ can be expanded into the infinitesimal fields 
\begin{equation}
    \bQh(\br,\tau)=\Sgh(\br,\tau)+\phi^i(\br,\tau)\bAh_i
    \label{eq:grass_pert}
\end{equation}
In terms of $\phi^i$, 
\begin{multline}
    S[c,\cb,\phi] =\int\rd\tau\left[
        \sum_{p_y}\cb_{a,p_y}\left(\delta^a{}_b\frac{\partial}{\partial\tau}-\lambda\Sgh^a{}_b\right)c^b_{p_y}\right.\\
        \left.-\lambda\sum_{p_y,\bq}F(p_y,\bq)\cb_{a,p_y+q_y/2}\phi^i(\Ah_i)^a{}_bc^b_{p_y-q_y/2}
    \right]
\end{multline}
Regarding the first term as the free action and the second term as the perturbation, the Feynman rules read\footnote{Here we adopt a shorthand notation that $\rp=(\omega,p_y)$ and $\rmq=(\nu,\bq)$.}
\begin{align}
    \begin{tikzpicture}[baseline=(bs.base),scale=0.7]
        \begin{feynhand}
            \vertex (bs) at (0,-0.1);
            \vertex (a) at (-1,0) {$a$};
            \vertex (b) at (1,0) {$b$};
            \propag[fer] (a) to [edge label=$\rp$] (b);
        \end{feynhand}
    \end{tikzpicture}&=G^a{}_b(\omega)\nonumber\\
    \bG(\omega)&=(i\omega-\lambda\Sgh)^{-1}\nonumber\\
    \begin{tikzpicture}[baseline=(bs.base),scale=0.7]
        \begin{feynhand}
            \vertex (bs) at (0,-0.1);
            \vertex[dot] (o) at (0,0) {};
            \vertex (a) at (-2,0) {$I$};
            \vertex (b) at (1,1.5) {$a$};
            \vertex (c) at (1,-1.5) {$b$};
            \propag[sca, mom=$\rmq$] (a) to (o);
            \propag[antfer] (o) to [edge label=$\rp-\half\rmq$] (b);
            \propag[antfer] (c) to [edge label=$\rp+\half\rmq$] (o);
        \end{feynhand}
    \end{tikzpicture}&=(\Gamma_i)^a{}_b(p_y,\bq)\nonumber\\
    \Gamma_i(p_y,\bq)&=F(p_y,\bq)\bAh_i
\end{align}

To obtain an effective action with respect to $\phi^i$, we integrate out $c,\cb$
\begin{equation}
    Z=\int\cD\cb\,\cD c\,\cD\phi\,e^{-S[c,\cb,\phi]}=\int\cD\phi\,e^{-S_\mathrm{eff}[\phi^i]}
\end{equation}
To the second order, $S_\mathrm{eff}$ contains the kinetic term in the NLSM
$
    \begin{tikzpicture}[baseline=(bs.base),scale=0.7]
        \begin{feynhand}
            \vertex (bs) at (0,-0.1);
            \vertex (o) at (-0.5,0) ;
            \vertex[dot]  (a) at (0,0) {};
            \vertex[dot]  (b) at (1,0) {};
            \vertex (c) at (1.5,0);
            \propag[sca] (a) to (o);
            \propag[sca] (b) to (c);
            \propag[fer] (a) to[in=135, out=45] (b);
            \propag[fer] (b) to[in=315, out=225] (a);
        \end{feynhand}
    \end{tikzpicture}
$; to the fourth order, it contains the box diagram that corresponds to the WZW term that we desire
\begin{equation}
    \begin{tikzpicture}[scale=0.7]
        \begin{feynhand}
            \vertex[dot] (a) at (-1,1)  {};
            \vertex[dot] (b) at (1,1)   {};
            \vertex[dot] (c) at (1,-1)  {};
            \vertex[dot] (d) at (-1,-1) {};
            \vertex (a1) at (-2,2)  {$i$};
            \vertex (b1) at (2,2)   {$j$};
            \vertex (c1) at (2,-2)  {$k$};
            \vertex (d1) at (-2,-2) {$l$};
            \propag[sca, mom=$\rmq_1$] (a1) to (a);
            \propag[sca, mom=$\rmq_2$] (b1) to (b);
            \propag[sca, mom=$\rmq_3$] (c1) to (c);
            \propag[sca, mom=$\rmq_4$] (d1) to (d);
            \propag[fermion] (a) to[edge label=$\rp_1$] (b);
            \propag[fermion] (b) to[edge label=$\rp_2$] (c);
            \propag[fermion] (c) to[edge label=$\rp_3$] (d);
            \propag[fermion] (d) to[edge label=$\rp_4$] (a);
        \end{feynhand}
    \end{tikzpicture}
    \label{eq:box_diag}
\end{equation}

To compare this box diagram term with the WZW term in the NLSM, we first express the WZW action in terms of the $\phi^i$ fields. In the NLSM, the matrix field $\bQ(\br,\tau)$ is extended into auxiliary fourth dimension $\bQt(\br,\tau,u)$ parametrised by $0\leq u\leq 1$ with boundary condition 
\begin{equation*}
    \bQt(\br,\tau,u=1)=\bQ(\br,\tau),\quad \bQt(\br,\tau,u=0)=\Sigma
\end{equation*}
Specifically, we take 
\begin{equation}
    \bQt(\br,\tau,u)=\Sigma+u\delta\bQ=\Sigma+u\phi^i\bA_i
\end{equation}
\begin{align}
    &S_\WZW\nonumber\\
    &=i\kappa\int_0^1\rd u\int\rd^2\br\,\rd\tau\,\epsilon^{\mu\nu\rho\sigma}\tr(\bQt\partial_\mu\bQt\partial_\nu\bQt\partial_\rho\bQt\partial_\sigma\bQt)\nonumber\\
    &=24i\kappa\int_0^1\rd u\int\rd^2\br\,\rd\tau\tr(\bQt\partial_x\bQt\partial_y\bQt\partial_\tau\bQt\partial_u\bQt)\nonumber\\
    &=24i\kappa\int_0^1u^3\rd u\int\rd^2\br\,\rd\tau\tr(\Sigma\partial_x\delta\bQ\partial_y\delta\bQ\partial_\tau\delta\bQ\delta\bQ)\nonumber\\
    &=6i\kappa\tr(\Sigma\bA_i\bA_j\bA_k\bA_l)\int\rd^2\br\,\rd\tau\partial_x\phi^i\partial_y\phi^j\partial_\tau\phi^k\phi^l    
\end{align}
where $\kappa$ is quantised to be 
\begin{equation*}
    \kappa=\frac{2\pi k}{(16\pi)^2}
\end{equation*}

On the other hand, the box diagram~\eqref{eq:box_diag} gives
\begin{widetext}
\begin{align}
    S_\Box&=-\int\frac{\rd\nu_1}{2\pi}\frac{\rd\nu_2}{2\pi}\frac{\rd\nu_3}{2\pi}\sum_{\bq_1\bq_2\bq_3}\phi^i(\bq_1,\nu_1)\phi^j(\bq_2,\nu_2)\phi^k(\bq_3,\nu_3)\phi^l(\bq_4,\nu_4)\nonumber\\
    &\qquad\times\frac{1}{V}\sum_{p_y}\int\frac{\rd\omega}{2\pi}\tr\left(\bG(\omega_1)\Gamma_i(p_{1y},\bq_1)\bG(\omega_2)\Gamma_j(p_{y,2},\bq_2)\bG(\omega_3)\Gamma_k(p_{y,3},\bq_3)\bG(\omega_4)\Gamma_l(p_{y,4},\bq_4)\right)\nonumber\\
    &=\int\frac{\rd\nu_1}{2\pi}\frac{\rd\nu_2}{2\pi}\frac{\rd\nu_3}{2\pi}\sum_{\bq_1\bq_2\bq_3}\cI^\Box_{ijkl}(\{\nu,\bq\})\phi^i(\bq_1,\nu_1)\phi^j(\bq_2,\nu_2)\phi^k(\bq_3,\nu_3)\phi^l(\bq_4,\nu_4)
\end{align}
\end{widetext}
where 
\begin{align*}
    \rmq_4&=-\rmq_1-\rmq_2-\rmq_3\\
    \rp_{1}&=\rp+\half \rmq_{1}\\
    \rp_{2}&=\rp+\rmq_{1}+\rmq_{2}\\
    \rp_{3}&=\rp+\rmq_{1}+\rmq_{2}+\half \rmq_{3}\\
    \rp_{4}&=\rp-\half \rmq_{1}
\end{align*}
and the form factor\footnote{Since the integral does not depend on $p_y$, the sum $\frac{1}{V}\sum_{p_y}$ will give only a factor $1/2\pi$.}
\begin{multline}
    \cI^\Box_{ijkl}(\{\nu,\bq\})=-\frac{1}{V}\sum_{p_y}\int\frac{\rd\omega}{2\pi}\\
    \times\tr\big(\bG(\omega_1)\Gamma_i(p_{1y},\bq_1)\bG(\omega_2)\Gamma_i(p_{y,2},\bq_2)\\
    \times\bG(\omega_3)\Gamma_i(p_{y,3},\bq_3)\bG(\omega_4)\Gamma_i(p_{y,4},\bq_4)\big)
    \label{eq:box_res}
\end{multline}

Writing the WZW term in the same form
\begin{multline*}
    S_\WZW=\int\frac{\rd\nu_1}{2\pi}\frac{\rd\nu_2}{2\pi}\frac{\rd\nu_3}{2\pi}\sum_{\bq_1\bq_2\bq_3}\cI^\WZW_{ijkl}(\{\nu,\bq\})\\
    \times\phi^i(\bq_1,\nu_1)\phi^j(\bq_2,\nu_2)\phi^k(\bq_3,\nu_3)\phi^l(\bq_4,\nu_4)
\end{multline*}
\begin{equation}
    \cI^\WZW_{ijkl}(\{\nu,\bq\})=6\kappa\tr(\Sigma\bA_i\bA_j\bA_k\bA_l)q_{1x}q_{2y}\nu_3
\end{equation}
We need to match $\cI^\Box$ and $\cI^\WZW$ in the long-wavelength limit $\{\bq,\nu\}\to0$
\begin{equation}
    6\kappa\tr(\Sigma\bA_i\bA_j\bA_k\bA_l)=\left.\frac{\partial}{\partial q_{1x}}\frac{\partial}{\partial q_{2y}}\frac{\partial}{\partial\nu_{3}}\cI^\Box_{ijkl}\right|_{\{\bq,\nu\}=0}
    \label{eq:box_comp}
\end{equation}

\subsection{Models}

\paragraph{Warm-up: $\SO(5)$ deconfined criticality} In the $\Sp(2)$ DQCP, the Grassmannian is parametrised by the $\Gamma$-matrices.
\begin{equation}
    \bQ=\bQh=n^i\Gamma_i,\quad n^in_i=1
\end{equation}
We choose one component as the reference and the rest as the perturbations
\begin{equation}
    \Sigma=\Gamma_0,\quad\mathbf{A}=(\Gamma_2,\Gamma_3,\Gamma_4,\Gamma_5)
\end{equation}
Under this choice, $\tr(\Sigma\bA_i\bA_j\bA_k\bA_l)=4\epsilon_{ijkl}$. By evaluating Eqs.~\eqref{eq:box_res} and \eqref{eq:box_comp}, we recover the result
\begin{equation*}
    \kappa=\frac{2\pi k}{(16\pi)^2},\quad k=1.
\end{equation*}

\paragraph{Symplectic Grassmannian $\rG=\Sp$} We now move on to a more general case of the Grassmannian
\begin{equation*}
    \frac{\Sp(N)}{\Sp(M)\times\Sp(N-M)}.
\end{equation*}
We consider $N_f=2N$ fermions in the fundamental representation. The generators in Eq.~\eqref{eq:lie_gen} are parametrised by 
\begin{align}
    (t_{(a_0b_0)})^a{}_b&=\half(\delta^a{}_{a_0}\Omega_{b_0b}+\delta^a{}_{b_0}\Omega_{a_0b})\nonumber\\
    (A_{(a_0b_0)})^a{}_b&=\delta^a{}_{a_0}\Omega_{b_0b}-\delta^a{}_{b_0}\Omega_{a_0b}\nonumber\\
    a_0&=1,\dots,2M,\ b_0=2M+1,\dots,2N
\end{align}
where the symplectic two form is taken as $\Omega=\left(\begin{smallmatrix}
    i\sigma^2&&\\
    &\ddots&\\
    &&i\sigma^2
\end{smallmatrix}\right)$, the index $i=(a_0b_0)$ denotes that $\bt_i$ lives in the symmetric rank-2 tensor representation, and the $A_{(a_0b_0)}$ outside the range of $a_0\leq 2M$ and $b_0>2M$ vanish. Correspondingly, the $\phi^i$ fields live in the bi-fundamental representation of $\Sp(M)\times\Sp(N-M)$. 
By evaluating Eqs.~\eqref{eq:box_res} and \eqref{eq:box_comp} component by component, we obtain
\begin{equation*}
    \kappa=\frac{2\pi}{(16\pi)^2},\quad k=1.
\end{equation*}

Several comments:
\begin{enumerate}
    \item The $\Sp(2)$ case is also recovered, as $\phi^i$ lives in the fundamental representation of $\mathrm{O}(4)=\Sp(1)\times\Sp(1)$;
    \item The theories with different $(N,M)$ have the same WZW level, as the components $\cI^\Box_{ijkl}$ for smaller $N$ and $N-M$ is a subset of larger ones.
\end{enumerate}

\paragraph{Orthogonal Grassmannian $\rG=\rO$} We consider $N_f=N$ fermions in the vector representation. The generators in Eq.~\eqref{eq:lie_gen} are parametrised by 
\begin{align}
    (t_{[a_0b_0]})_{ab}&=\half(\delta_{aa_0}\delta_{bb_0}-\delta_{ab_0}\delta_{ba_0})\nonumber\\
    (A_{[a_0b_0]})_{ab}&=\delta_{aa_0}\delta_{bb_0}+\delta_{ab_0}\delta_{ba_0}\nonumber\\
    a_0&=1,\dots,M,\ b_0=M+1,\dots,N
\end{align}
By evaluating Eqs.~\eqref{eq:box_res} and \eqref{eq:box_comp} component by component, we obtain
\begin{equation*}
    \kappa=\frac{2\pi}{(16\pi)^2}.
\end{equation*}
Note that this now correspond to 
\begin{equation*}
    k=2,
\end{equation*}
because for $\rO$ and $\SO$ Grassmannian, the quantisation of the WZW term is halved.

\paragraph{Unitary Grassmannian $\rG=\rU$} We consider $N_f=N$ fermions in the fundamental representation. The generators in Eq.~\eqref{eq:lie_gen} are parametrised by 
\begin{align}
    (t_{a_0}{}^{b_0})^a{}_b&=\delta^a_{a_0}\delta^{b_0}_b\nonumber\\
    (A_{a_0}{}^{b_0})^a{}_b&=\left\{\begin{aligned}
        &2\delta^a_{a_0}\delta^{b_0}_b,&a_0&=1,\dots,M,\\
        &&b_0&=M+1,\dots,N\\
        -&2\delta^a_{a_0}\delta^{b_0}_b,&a_0&=M+1,\dots,N,\\
        &&b_0&=1,\dots,M
    \end{aligned}\right.
\end{align}
By evaluating Eqs.~\eqref{eq:box_res} and \eqref{eq:box_comp} component by component, we obtain
\begin{equation*}
    \kappa=\frac{2\pi}{(16\pi)^2},\quad k=1.
\end{equation*}

We note that the $\rU$ Grassmannian possesses an additional $\rU(1)_m$ symmetry associated with the non-trivial homotopy group $\pi_2=\mathbb{Z}$. In a CFT arising from the corresponding NLSM, this $\rU(1)$ symmetry cannot be identified with the electric $\rU(1)_e$, since $\rU(1)_e$ is necessarily gapped on the fuzzy sphere. 

\paragraph{Manipulations of the WZW level} We now consider more general cases beyond $N$ flavours of fermions in the fundamental representation of $\rG(N)$ and $\bQh=\bQ$. We discuss how different manipulations can modify the WZW level. 
\begin{enumerate}
    \item Complex conjugate does not change the WZW level:
    \begin{align*}
        \bQh'&=\bQh^\dagger,&\bth'&=-\bth^\dagger,&\bAh'=&\bAh^\dagger\\\cI'^\Box&=\cI^\Box,&\kappa'&=\kappa
    \end{align*}
    \item Particle-hole transformation inverses the WZW level:
    \begin{equation*}
        \Sgh'=-\Sgh,\quad\bAh'=-\bAh,\quad\cI'^\Box=-\cI^\Box,\quad\kappa'=-\kappa
    \end{equation*}
    \item Stacking sums over the WZW level:
    \begin{equation*}
        \bQh'=\bQh_1\oplus\bQh_2,\quad\cI'^\Box=\cI^\Box_1+\cI^\Box_2,\quad\kappa'=\kappa_1+\kappa_2
    \end{equation*}
\end{enumerate}

\subsection{Unitary Grassmannian: Matching the Magnetic $\rU(1)_m$}

In the $\rG=\rU$ Grassmannian, there is an additional $\rU(1)_m$ symmetry. To realise a CFT on the LLL, we should identify both the $\SU(N)$ and the $\rU(1)_m$ as sub-groups of the maximal $\SU(N_f)$ global symmetry. To see if the $\rU(1)_m$ has the correct winding with $\SU(N)$, we couple the $\rU(1)_m$ current to a gauge field $a_\mu$ through minimal coupling.
\begin{gather}
    Z=\int\cD\psi\,\cD\psb\,\cD\bQh\,\cD a\,e^{-S_\psi-S_Q-S_a}\nonumber\\
    S_a[\psi,\psb,a]=-\int\rd^2\br\,\rd\tau\,a_\mu j^\mu\nonumber\\
    j^0=\psb\bth_m\psi,\qquad j^{1,2}=0\nonumber\\
    \begin{tikzpicture}[baseline=(bs.base),scale=0.7]
        \begin{feynhand}
            \vertex (bs) at (0,-0.1);
            \vertex[dot] (o) at (0,0) {};
            \vertex (a) at (-2,0) {$\mu$};
            \vertex (b) at (1,1.5) {$a$};
            \vertex (c) at (1,-1.5) {$b$};
            \propag[bos, mom=$\rmq$] (a) to (o);
            \propag[antfer] (o) to [edge label=$\rp-\half\rmq$] (b);
            \propag[antfer] (c) to [edge label=$\rp+\half\rmq$] (o);
        \end{feynhand}
    \end{tikzpicture}=\delta^\mu{}_0T^a{}_bF(p_y,\bq)
\end{gather}
where $\bth_m$ is the $\rU(1)_m$ generator.\footnote{The fermions only couple to the temporal component of the gauge field, we expect the Lorentz symmetry to emerge.} Integrating out the fermions, the gauge field is coupled to the matrix field in the third order through the triangle diagram 
\begin{equation}
    \begin{tikzpicture}[scale=0.7]
        \begin{feynhand}
            \vertex[dot] (a) at (-1,-1) {};
            \vertex[dot] (b) at (-1,1)  {};
            \vertex[dot] (c) at (1,0)   {};
            \vertex (a1) at (-2,-2) {$i$};
            \vertex (b1) at (-2,2)  {$j$};
            \vertex (c1) at (2.5,0) {$\mu$};
            \propag[sca, mom=$\rmq_1$] (a1) to (a);
            \propag[sca, mom=$\rmq_2$] (b1) to (b);
            \propag[bos, mom=$\rmq_a$] (c1) to (c);
            \propag[fermion] (a) to[edge label=$\rp_1$] (b);
            \propag[fermion] (b) to[edge label=$\rp_2$] (c);
            \propag[fermion] (c) to[edge label=$\rp_3$] (a);
        \end{feynhand}
    \end{tikzpicture}
    \label{eq:tri_diag}
\end{equation}

In the NLSM, this corresponds to coupling the gauge field $a_\mu$ to the Skyrmion current 
\begin{align}
    S_\Skyr&=\frac{q_m}{2\pi}\int\rd^2\br\,\rd\tau\,\epsilon^{\mu\nu\rho}a_\mu\tr(\bQ\partial_\nu\bQ\partial_\rho\bQ)\nonumber\\
    &=\frac{q_m}{2\pi}\tr(\Sigma\bA_i\bA_j)\int\rd^2\br\,\rd\tau\,\epsilon^{\mu\nu\rho}a_\mu\partial_\nu\phi^i\partial_\rho\phi^j\nonumber\\
    &=\int\frac{\rd\nu_1}{2\pi}\frac{\rd\nu_2}{2\pi}\sum_{\bq_1\bq_2}\cI^{\mu,\Skyr}_{ij}\nonumber\\
    &\qquad\times a_\mu(\bq_a,\nu_a)\phi^i(\bq_1,\nu_1)\phi^j(\bq_2,\nu_2)\nonumber\\
    \cI^{\mu,\Skyr}_{ij}&=\frac{q_m}{2\pi}\tr(\Sigma\bA_i\bA_j)\epsilon^{\mu\nu\rho}q_{1\nu}q_{2\rho}
\end{align}
where $q_m$ is the Skyrmion charge. Evaluating the triangle diagram Eq.~\eqref{eq:tri_diag} explicitly,
\begin{widetext}
\begin{align}
    S_{\triangle}&=-\int\frac{\rd\nu_1}{2\pi}\frac{\rd\nu_2}{2\pi}\sum_{\bq_1\bq_2}a_0(\bq_a,\nu_a)\phi^i(\bq_1,\nu_1)\phi^j(\bq_2,\nu_2)\nonumber\\
    &\qquad\times\frac{1}{V}\sum_{p_y}\int\frac{\rd\omega}{2\pi}\tr\left(\bG(\omega_1)\Gamma_i(p_{1y},\bq_1)\bG(\omega_2)\Gamma_i(p_{2y},\bq_2)\bG(\omega_3)\bth_m F(p_{3y},\bq_3)\right)\nonumber\\
    &=\int\frac{\rd\nu_1}{2\pi}\frac{\rd\nu_2}{2\pi}\sum_{\bq_1\bq_2}\cI^{0,\triangle}_{ij}(\{\nu,\bq\})a_0(\bq_a,\nu_a)\phi^i(\bq_1,\nu_1)\phi^j(\bq_2,\nu_2)\nonumber\\
    \cI_{ij}^{\mu,\triangle}&=\frac{1}{V}\sum_{p_y}\int\frac{\rd\omega}{2\pi}\tr\left(\bG(\omega_1)\Gamma_i(p_{1y},\bq_1)\bG(\omega_2)\Gamma_i(p_{2y},\bq_2)\bG(\omega_3)\bth_m F(p_{3y},\bq_3)\right)
\end{align}
\end{widetext}
Matching $\cI^{\triangle}$ and $\cI^\Skyr$ in the long-wavelength limit, we obtain the Skyrmion charge of $\rU(1)_m$.
\begin{equation}
    \frac{q_m}{2\pi}\tr(\Sigma\bA_i\bA_j)=\left.\frac{\partial}{\partial q_{1x}}\frac{\partial}{\partial q_{2y}}\cI^{0,\triangle}_{ij}\right|_{\{\bq,\nu\}=0}
\end{equation}

\section{QMC Details}
\label{app:QMC_details}

As briefly mentioned in the main text, we are using a higher-order Trotter decomposition\cite{Goth2022}. Let us assume $H=\sum_{i=1}^N O_i$ to simplify the notation, thus
\begin{align}
    Z &= \Tr{\big[e^{-\beta H}\big]}\nonumber\\
      &= \Tr{\Bigg[\prod_{\tau=1}^{L_{\text{trot}}}\prod_{s=1}^2 U_{\tau,s}\Bigg]} +\mathcal{O}(\Delta\tau^2)\quad\mathrm{with,}\nonumber\\
    U_{\tau,s} &= e^{d_s\Delta\tau O_N} \dots e^{d_s\Delta\tau O_1}e^{c_s\Delta\tau O_1}\dots e^{c_s\Delta\tau O_N}\,.
\end{align}
The coefficients are $c_1=d_2=0.21178$ and $c_2=d_1=0.28822$. Similar to $e^{A+B}\approx e^{A/2}e^Be^{A/2}$, this decomposition is symmetric, but the $c_s$ and $d_s$ coefficients have been tuned to minimize the prefactor of the leading order Trotter error \cite{Goth2022}.

Furthermore, we scale the inverse temperature $\beta =10 s$ with the monopole flux $s$, i.e., the system size, such that all presented results are ground state properties. We use $\Delta\tau=0.1$ throughout the manuscript. 

\begin{figure}[htbp]
    \centering
    \includegraphics[width=.35\textwidth]{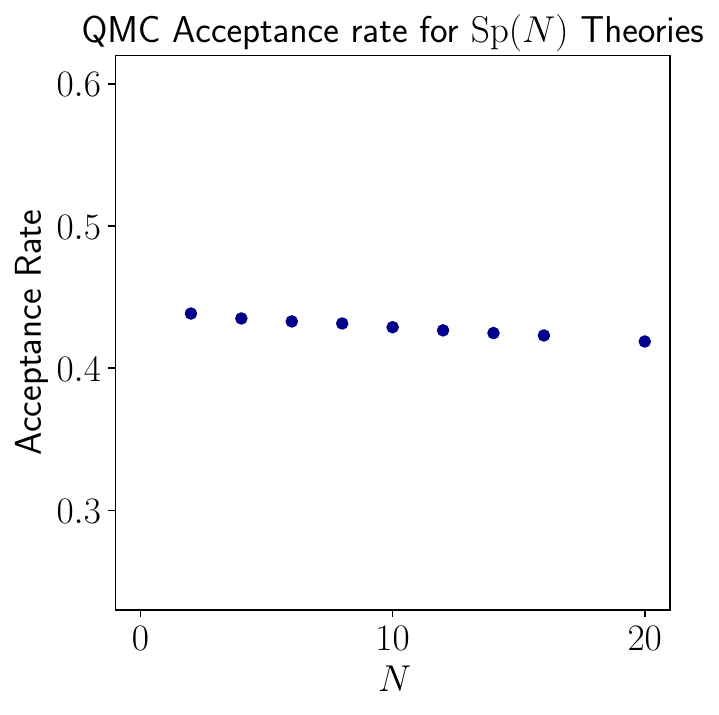}
    \caption{QMC acceptance rate as a function of $N$ for $\mathrm{Sp}(N)$ theories with $N_{\mathrm{orb}}=15$. There is very little reduction in acceptance even for $N=20$. Error bars are smaller than the point size.} 
    \label{fig:acceptance}
\end{figure}

In Figure \ref{fig:acceptance}, we plot the average acceptance rate of proposed auxiliary field updates with respect to the number of flavours $N$ in the $\mathrm{Sp}(N)$ theories at a constant $N_{\mathrm{orb}}=15$. The acceptance rate roughly stays around $40-50\%$ and only slightly decreases with increasing $N$. Hence, this algorithm is well-suited to studying large-$N$ theories, retaining the polynomial scaling of sign-problem-free QMC.

\begin{figure}[htbp]
    \centering
    \includegraphics[width=.48\textwidth]{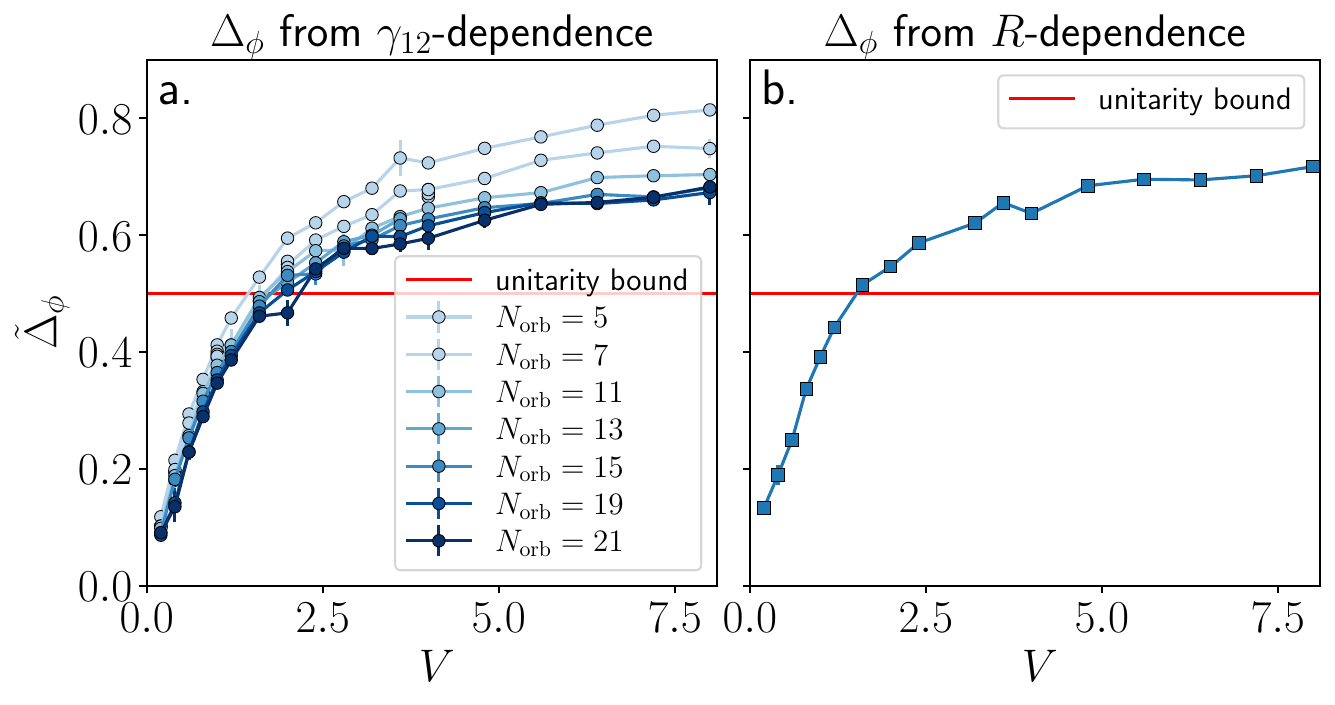}
    \caption{Extrapolated $\tilde{\Delta}_\phi$ by $\gamma_{12}$-dependence (a), and $R$-dependence (b). In contrast to the $\mathrm{Sp}(4)$ and $\mathrm{Sp}(10)$ theories, for $\mathrm{Sp}(2)$ we do not see a crossing point.} 
    \label{scalingcollapses2}
\end{figure}

\begin{figure*}
    \includegraphics[width=.35\textwidth]{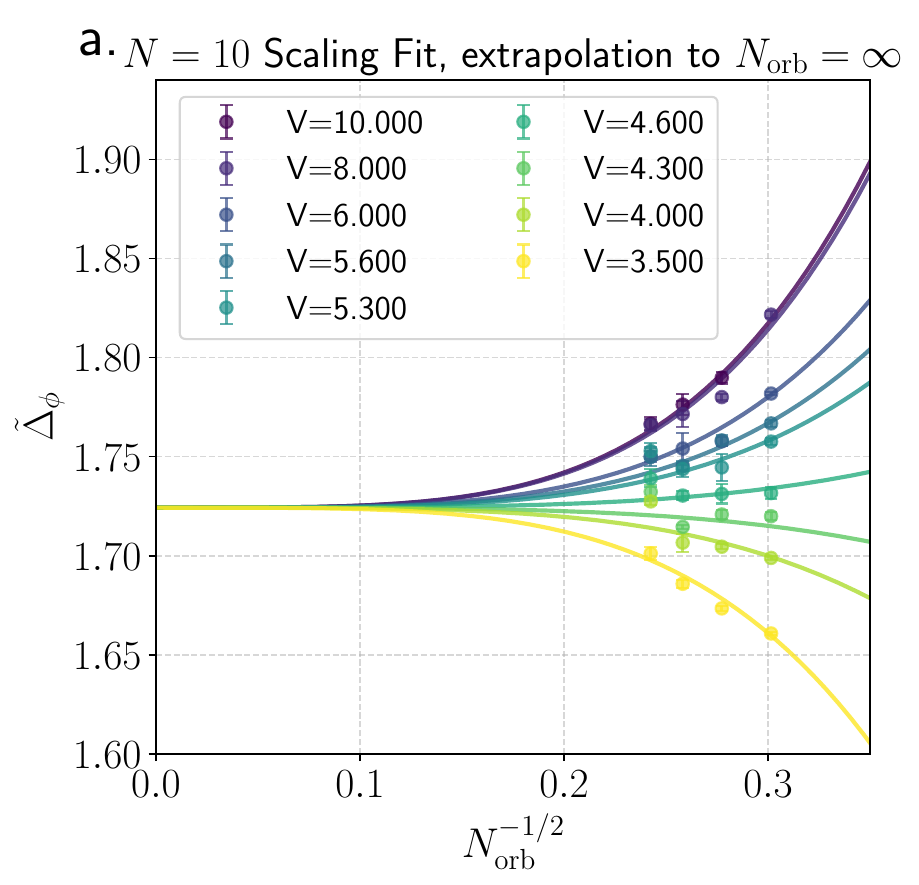}
    \includegraphics[width=.35\textwidth]{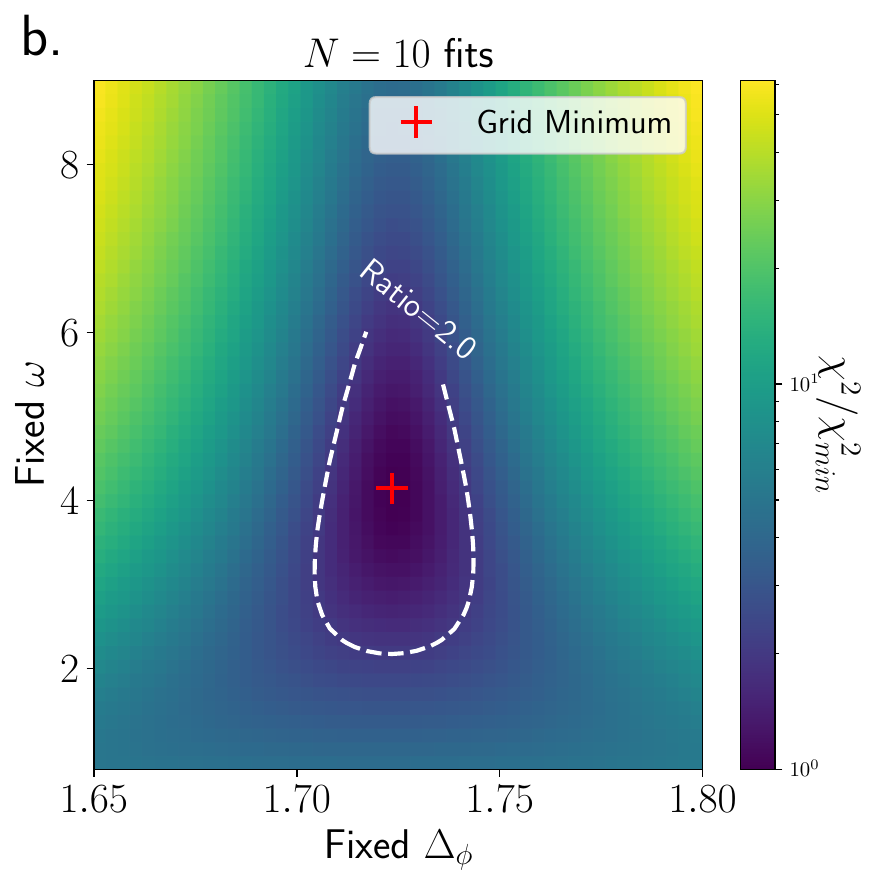}
    \caption{Fits to determine $\Delta_\phi$ for $N=10$. The plot to the left shows data that is fit using the ansatz of $\Delta_\phi(V,N_{\mathrm{orb}}) = \Delta_\phi + g(V)N_{\mathrm{orb}}^{-\omega/2} $. The observable used for $\Delta_\phi(V,N_{\mathrm{orb}})$ is the RG-invariant $\partial \cC_A\partial\gamma_{12} $ at $\gamma_{12}=7\pi/9$. The plot to the right gives the chi-square values for fits relative to the best chi-square fit, for a grid of fixed $\Delta$ and $\omega$ values.
    }
    \label{fit4}
\end{figure*}

\section{QMC Data for \texorpdfstring{$\tilde{\Delta}_\phi$}{Delta phi} and Fitting Details}
\label{app:fit_dim}
First, we present the $\tilde{\Delta}_\phi$ results for the model with $\Sp(2)$, relevant for $\SO(5)$ deconfined quantum criticality, in analogy to Figure~\ref{conformalform4} of the main text. In doing so we benchmark the QMC simulations against the exact diagonalization results in \cite{Zhou2023SO5}. In Figure \ref{scalingcollapses2}a we show $\tilde{\Delta}_\phi$ from
\begin{equation}
\begin{aligned}
\tilde{\Delta}_\phi(\theta) =& -\tan\gamma_{12}\left.\frac{\partial \log \cC_A}{\partial \gamma_{12}}\right|_{\gamma_{12}=7\pi/9},
\label{deltaphi}
\end{aligned}
\end{equation}
as a function of $V$ and $N_{\mathrm{orb}}$. The value of $\gamma_{12}=7\pi/9$ is the same value used in the main text due to its statistical precision and reduced finite-size effects. While these plots in general use the same analysis as those in the main text for $\mathrm{Sp}(4)$ and $\mathrm{Sp}(10)$, they do not show a crossing in \ref{scalingcollapses2}a. Instead, the data systematically drifts towards smaller values with system size, approaching the unitarity bound. This is consistent with the scenario depicted in Figure \ref{fig:ph_diag}, i.e., pseudocriticality ($2<N_c$), for $\mathrm{Sp}(2)$, as found in \cite{Zhou2023SO5}.
Figure \ref{scalingcollapses2}b depicts the same quantity, $\tilde{\Delta}_\phi$, extracted via its size dependence, which is consistent with the data in panel a.

Finally, let us discuss the procedure used to determine `the' operator scaling dimension $\Delta_\phi$ for each $N$, as reported in the table of Figure~\ref{largenplot} of the main text.
Here, we consider the $\tilde{\Delta}_\phi$ data for a whole range of values of $N_{\mathrm{orb}}$ and $V$, which are roughly in the critical phase. For example, we consider the range $3.5 \leq V \leq 10$ for $\Sp(10)$. 
Since the lowest irrelevant operator dominates the RG-flow towards the stable CFT fixed point, the scaling corrections in its vicinity are given by~\cite{Zhou2025Jul}
\begin{equation}
\tilde{\Delta}_\phi(V,N_{\mathrm{orb}}) = \Delta_\phi + g(V)N_{\mathrm{orb}}^{-\omega/2}.
\end{equation}
Hence, we can fit the whole dataset and extract `the' scaling dimension $\Delta_\phi$ for each $\Sp(N)$ symmetry group. An exemplary fit is depicted in Figure~\ref{fit4}a. We evaluate the standard deviation, $\chi^2$ as a function of the fitting parameters $\Delta_\phi$ and $\omega$, shown in Figure~\ref{fit4}b and use the ratio $\chi^2/\chi^2_{\min}=2$ to estimate the confidence interval. For $\Sp(10)$, we get $\Delta_\phi=1.75(2)$.
A similar analysis is repeated for all other values of $N$ and the extracted scaling dimensions are summarized in Figure~\ref{largenplot} of the main text.

\FloatBarrier

\bibliography{ref}
\end{document}